\documentclass[journal]{IEEEtran}

\usepackage{acronym}
\usepackage{graphicx}
\usepackage[ruled,vlined]{algorithm2e}
\usepackage{multicol}
\usepackage[english]{babel}
\usepackage{amsmath,bm}
\usepackage[table,xcdraw]{xcolor}
\usepackage{mathtools}

\usepackage{setspace}

\usepackage[section]{placeins}

\usepackage{multirow}
    \usepackage[position=top]{subfig}
    \usepackage{floatrow}   
    
\usepackage{hyperref}
\hypersetup{%
    pdfborder = {0 0 0}
}

\acrodef{APS}[APS]{Active Pixel Sensor}
\acrodef{ATIS}[ATIS]{Asynchronous Time-Based Imaging System}
\acrodef{CCD}[CCD]{Charge-Coupled Device}
\acrodef{COM}[COM]{Centre of Mass}
\acrodef{CMOS}[CMOS]{Complementary Metal–Oxide Semiconductor}
\acrodef{DBSCAN}[DBSCAN]{Density-Based Spatial Clustering of Applications with Noise}
\acrodef{DVS}[DVS]{Dynamic Vision Sensor}
\acrodef{EB}[EB]{Event-Based}
\acrodef{EBC}[EBC]{Event-Based Camera}
\acrodef{EBS}[EBS]{Event-Based Sensing}
\acrodef{ETT}[ETT]{Extended Target Tracking}
\acrodef{EPS}[EPS]{Events per Second}
\acrodef{FAN}[FAN]{Fast Adapting Network}
\acrodef{FEAST}[FEAST]{Feature Extraction using Adaptive Selection Thresholds}
\acrodef{FIESTA}[FIESTA]{Fast Iterative Extraction of Salient targets for Tracking Asynchronously}
\acrodef{FISST}[FISST]{Finite Set Statistics}
\acrodef{FOR}[FOR]{Field of Regard}
\acrodef{FOV}[FOV]{Field-of-View}
\acrodef{FPS}[FPS]{Frames Per Second}
\acrodef{GEO}[GEO]{Geostationary Earth Orbit}
\acrodef{GNN}[GNN]{Global Nearest Neighbour}
\acrodef{GOSPA}[GOSPA]{Generalized Optimal SubPattern Assignment}
\acrodef{GSM}[GSM]{Gaussian Sum Filter}
\acrodef{ICRS}[ICRS]{International Celestial Reference System}
\acrodef{JPDA}[JPDA]{Joint Probabilistic Data Association}
\acrodef{LIF}[LIF]{Leaky Integrate and Fire}
\acrodef{MEO}[MEO]{Medium Earth Orbit}
\acrodef{MHT}[MHT]{Multiple Hypothesis Tracker}
\acrodef{MPMI}[MPMI]{Multiple Point Multiple Interval}
\acrodef{MTT}[MTT]{Multiple Target Tracking}
\acrodef{NN}[NN]{Nearest Neighbour}
\acrodef{OD}[OD]{Orbit Determination}
\acrodef{PDA}[PDA]{Probabilistic Data Association}
\acrodef{PDF}[PDF]{Probability Density Function}
\acrodef{PHD}[PHD]{Probability Hypothesis Density}
\acrodef{PMHT}[PMHT]{Probabilistic Multiple Hypothesis Tracker}
\acrodef{GM-PHD}[GM-PHD]{Gaussian Mixture Probability Hypothesis Density}
\acrodef{PMBM}[PMBM]{Poisson Multi-Bernoulli Mixture}
\acrodef{PMF}[PMF]{Probability Mass Function}
\acrodef{PSF}[PSF]{Point Spread Function}
\acrodef{LEO}[LEO]{Low Earth Orbit}
\acrodef{RFS}[RFS]{random finite set}
\acrodef{RH}[RH]{Riccardi-Honders}
\acrodef{RMSE}[RMSE]{Root Mean Squared Error}
\acrodef{ANEES}[ANEES]{Average Normalised Estimation Error Squared}
\acrodef{RANSAC}[RANSAC]{Random Sample Consensus}
\acrodef{RSO}[RSO]{Resident Space Object}
\acrodef{SAE}[SAE]{Surface of Activated Events}
\acrodef{SAN}[SAN]{Slow Adapting Network}
\acrodef{SPSI}[SPSI]{Single Point Single Interval}
\acrodef{SNR}[SNR]{Signal-to-Noise Ratio}
\acrodef{SSA}[SSA]{Space Situational Awareness}
\acrodef{STT}[STT]{Single Target Tracking}
\acrodef{STDP}[STDP]{Spike Timing Dependent Plasticity}
\acrodef{STM}[STM]{Space Traffic Management}
\acrodef{SWaP}[SWaP]{Size, Weight and Power}
\acrodef{TLE}[TLE]{Two-Line Element Set}
\acrodef{TTA}[TTA]{Target Time to Acquire}
\acrodef{WCS}{World Coordinate System}

\begin{document}
%
\title{Astrometric Calibration and Source Characterisation of the Latest Generation Neuromorphic Event-based Cameras for Space Imaging}
%
%
%

\author{Nicholas~Owen~Ralph,
        Alexandre~Marcireau,
        Saeed~Afshar,
        Nicholas~Tothill,
        Andr\'e~van~Schaik,
        and~Gregory~Cohen
\thanks{NO. Ralph, A. Marcireau, S. Afshar, N. Tothill, A. van Schaik, and G. Cohen are with the International Centre for Neuromorphic Systems, MARCS Institute for Brain Behaviour and Development, Western Sydney University.}
}

%
%

\markboth{Event-Based Star Mapping, Calibration and Characterisation November 2022}%
{Shell \MakeLowercase{\textit{et al.}}: Bare Demo of IEEEtran.cls for IEEE Journals}
%



\maketitle

\begin{abstract}


As an emerging approach to space situational awareness and space imaging, the practical use of an event-based camera in space imaging for precise source analysis is still in its infancy. The nature of event-based space imaging and data collection needs to be further explored to develop more effective event-based space image systems and advance the capabilities of event-based tracking systems with improved target measurement models. Moreover, for event measurements to be meaningful, a framework must be investigated for event-based camera calibration to project events from pixel array coordinates in the image plane to coordinates in a target resident space object's reference frame. In this paper, the traditional techniques of conventional astronomy are reconsidered to properly utilise the event-based camera for space imaging and space situational awareness. This paper presents the techniques and systems used for calibrating an event-based camera for reliable and accurate measurement acquisition. These techniques are vital in building event-based space imaging systems capable of real-world space situational awareness tasks. By calibrating sources detected using the event-based camera, the spatio-temporal characteristics of detected sources or `event sources' can be related to the photometric characteristics of the underlying astrophysical objects. Finally, these characteristics are analysed to establish a foundation for principled processing and observing techniques which appropriately exploit the capabilities of the event-based camera. 

\end{abstract}


%
\IEEEpeerreviewmaketitle


\section{Introduction}\label{sect: measuring_the_sky_intro}

\IEEEPARstart{E}{arth} orbit is a vantage point for many commercial, scientific, and defence satellites. High-demand orbital slots are rapidly populating, increasing the likelihood of collision between these \acp{RSO} \cite{liou2006risks}. The proliferation of Earth orbit requires a sound approach to \acl{SSA} to mitigate these hazards by gathering information on the environment of space to coordinate satellite operations. Optical space imaging approaches for \ac{SSA} are common for the observation of extra-atmospheric objects such as satellites and astrophysical objects \cite{bobrinsky2010space}. Recently, the \ac{EBC} as a novel class of imaging device, has been demonstrated as an alternative and new approach in space imaging and \ac{SSA} without many of the constraints of typical frame-based imaging sensors such as \acp{CCD} and \acp{CMOS}-based \acp{APS} \cite{cohen2019event}. 

The \acl{EBC} is a biology-inspired class of vision sensor that draw inspiration from the photo-receptors in biological eyes. This inspiration is based on the mechanisms that provide the retina with its impressive capabilities \cite{mahowald1994silicon}, namely how the retina functions to extract and efficiently encode meaningful information from the visual scene \cite{mead2020we}. Each pixel in an \ac{EBC} contains analogue circuitry that operates as a contrast detector. Each pixel is also independent and asynchronous from one another and only emit data when either a positive or negative contrast change is detected.

\begin{figure}[!ht]
    \centering
    \includegraphics[width=\textwidth]{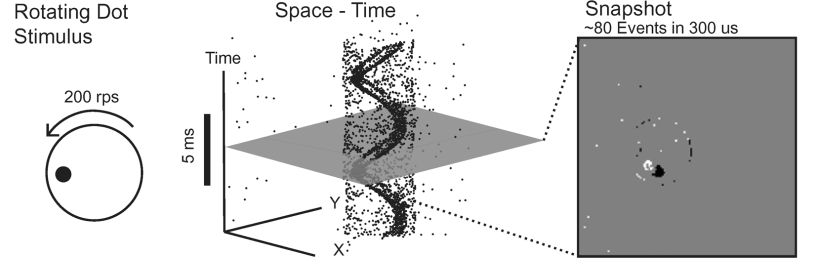}
        \caption{The response of an \ac{EBC} to a rotating dot stimulus, which produces a continuous stream of events with high temporal resolution. Using a frame-based imaging system, discrete image frames would instead be produced in discrete sample intervals which are more prone to blur effects. From \cite{lichtsteiner2008128}.}
        
    \label{fig:ebc_comparison}

\end{figure}{}

\begin{figure*}[]
    \centering
    \includegraphics[width=\textwidth]{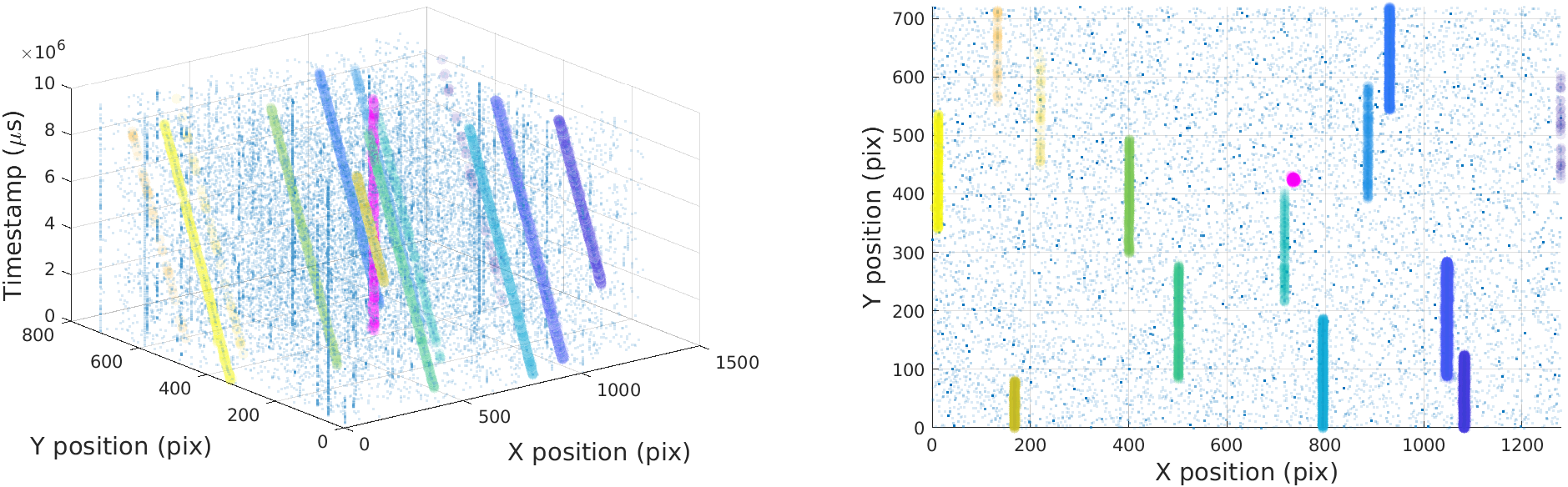}
    \caption{Labelled 10-second Gen 4 HD \ac{ATIS} observation of MEO BEIDOU-3 M1 (NORAD ID 43001) (magenta) in the 3D spatio-temporal event-stream (left) and a 2D event accumulation (right) using standard mount tracking (right). Background stars appear among the background noise as streaks moving through the FOV (colour overlay given by target label).}
    \label{fig:demo_2dvs3d_rso_observation_chapter2}
\end{figure*}

As these sensors detect change in contrast rather than integrating the light at each pixel, the pixels have an extremely high dynamic range. Furthermore, the independent nature of each pixel allows them to look for changes around their own individual set points, allowing them to avoid the problems around saturation suffered by fixed-time integrating cameras \cite{gallego2019event}. This pixel-level change also causes these sensors to produce data that is fundamentally different from the output of a conventional camera \cite{chen2020event}, as shown in Figure \ref{fig:ebc_comparison}. In place of frames, these devices output a stream of change events with microsecond resolution. The output of these sensors is also activity-dependent, which allows them to produce far less data than a conventional camera for sparse scenes. Combined with the high dynamic range, this makes these sensors extremely well suited for space imaging applications. The typical event-stream output of an \ac{RSO} observation using the latest Gen 4 HD \ac{ATIS} is illustrated in Figure \ref{fig:demo_2dvs3d_rso_observation_chapter2}. 

The \ac{EBC} has been successfully demonstrated as a new approach to \ac{SSA} and space imaging \cite{cohen2019event}. In this first study, an \ac{EB} telescopic space imaging system was demonstrated to be capable of observing \acp{RSO} in \ac{LEO} and \ac{GEO}. In these experiments, the \ac{EB} space imaging system performed with a significantly higher temporal resolution and lower output data volume than conventional \acp{CCD} sensors, which require longer exposure times and suffer from blur effects when observing moving targets.

Dedicated \ac{RSO} tracking algorithms for \ac{EB} vision data have been established in the literature using frame-based \cite{cheung2018probabilistic} and event-based \cite{afshar2020event} representation and processing techniques. \acp{RSO} and aircraft have been observed using \acp{EBC} in various conditions and optical setups \cite{ralph2019observations}. Frame-based star trackers to transform \ac{EB} pixel measurements to a world coordinate frame using astrometric calibration have also been developed. This so-called `star mapping' has been performed using integrated event-frames with manual motion compensation \cite{cohen2018approaches}, frame generation using a basic form of contrast maximisation \cite{chin2019star}, and motion compensation using Hough line detection \cite{bagchi2020event}. \ac{EB} datasets have been developed in these articles with hand-labelled real-world tracking data \cite{cheung2018probabilistic, afshar2020event} and physically simulated astrophysical sources \cite{chin2019star, bagchi2020event}. Space-based operation of the \ac{EBC} for spacecraft landing on the Moon has been investigated using simulated data \cite{sikorski2021event}. The use of \acp{EBC} for satellite characterisation has been also studied \cite{jolley2022evaluation}. The viability of \ac{EBC} operation in the high-radiation environment of \ac{LEO} has been tested using a physical simulation \cite{roffe2021neutron}, where the authors conclude that the \ac{EBC} will be robust to on-orbit operation in the presence of neutron-radiation.

The viability of the \ac{EBC} has been assessed in \cite{zolnowski2019observational} and \cite{mcmahon2021commercial}. These articles discuss the limitations and advantages of the \ac{EBC} and conclude that although the \ac{EBC} is less sensitive than comparative integrating \ac{CCD} sensors, the \ac{EBC} represents a new potential avenue for space imaging, given the inherent high time resolution, low latency, and low volume data output.  \cite{mcmahon2021commercial} in particular evaluate the \ac{EBC} and perform a large sky survey using a third generation \ac{EBC} (more detail on \ac{EBC} generations discussed in Section \ref{tab:chapter_2_EBC_specs}). In this work, the authors establish a model for the limiting magnitude of the third generation \acp{EBC}, which are empirically evaluated by observing stars while sidereally tracking. The authors also evaluate the limiting magnitude of the \ac{EBC} at varying angular scan rates. This work concludes that the \ac{EBC} shows clear promise for small-aperture, large-field persistent \ac{SSA} in terms of their ``efficient capture of temporal information''. A model of the \ac{EBC} is also developed \cite{savransky2021rachel} for simulating \ac{EB} data for \ac{SSA} in which a physics-based end-to-end model for event-based sensing of \acp{RSO} is proposed. This model was designed to serve as a tool to advise system design decisions for \ac{SSA} and to provide a consistent simulated output for algorithm development with \acp{EBC}. A similar simulator proposed in \cite{joubert2021event} generates simulated \ac{RSO} observations for the purposes of developing and evaluating an \ac{EB} tracking algorithm for \ac{SSA}. These works were based on third generation \acp{EBC}, however, recently, the Prophesee `Gen 4 HD' \ac{EBC} has been made available with significant improvements to performance, as detailed in Table \ref{tab:chapter_2_EBC_specs}.

The overall goal of \ac{SSA} operators is clear; to provide fast, accurate and reliable information on \acp{RSO}. The characteristics of \ac{EB} space imaging systems are well suited to these goals. The high temporal resolution and low latency of the \ac{EBC} allow an \ac{EB} imaging system to rapidly observe targets, in addition to responding quickly to new and unexpected \acp{RSO}. The low output data volume produced by \ac{EBC} is attractive to \ac{SSA} operators by reducing the power, storage, and cost requirements of traditional ground-based and space-based \ac{SSA} networks. Easing these requirements and facilitating an increased \ac{RSO} track capacity is vital as data collection efforts increase in response to the rapidly growing \ac{RSO} population. 

Event-based sensing systems represent a potentially disruptive technology for conventional imaging and particularly space imaging. Few studies have been conducted on the observation limits of the \ac{EBC}, the spatio-temporal characteristics of event sources, their relationship with target photometric properties, and how these characteristics change while observing at varying angular mount scan speeds. Night sky limiting magnitudes have been independently verified as magnitude 10.38 in \cite{zolnowski2019observational} using the third generation \ac{DVS} with 320~ms duration event accumulation frames at 1$\times$ sidereal rate, and as faint as magnitude 5.45 while performing a scanning slew at 300$\times$ times sidereal rate. Similar limits were observed in \cite{cohen2019event} using the same sensors at magnitude 10.19 at 1$\times$ sidereal tracking with smaller aperture telescopes. The current most detailed study of the spatio-temporal characteristics of source observation is from \cite{mcmahon2021commercial}. The authors establish a model for the limiting magnitude of the third generation \ac{ATIS} and \ac{DVS} \acp{EBC}, which are empirically evaluated by observing stars while sidereally tracking. The authors propose a theoretical limit of 9.6 in these evaluations and observe a limiting magnitude of 9.8 using relatively small aperture telescopes. Also evaluated was the limiting magnitude of the \ac{EBC} at varying angular scan rates, reporting that no sources could be detected at 0.5 degrees per second. The authors noted that the empirical limiting magnitude dropped by approximately 1.5 magnitudes as the scan rate increased from 0.01 to 0.1 degrees per second. Additionally, it was shown that the event rate of a source increases rapidly as the apparent brightness increases.


While the \ac{EBC} has demonstrated clear advantages as a new and complementary approach to \ac{SSA} and space imaging, the development of algorithms to process this novel data is difficult. To yield functional \ac{RSO} state estimates using \ac{EB} data, precise position and timing information must be gathered from a space imaging system to convert event measurements in the pixel frame to a standard world coordinate frame. To date, a real-time and online \ac{EB} \ac{RSO} tracking system that sufficiently leverages the capabilities of the \ac{EBC} and can produce calibrated, accurate, and precise state estimations of \ac{RSO} has not yet been published in the literature. The in-depth analysis of the relationships between spatio-temporal features and the photometric characteristics of faint point sources and bright extended sources at varying scan rates and different seeing conditions is yet to be produced. Additionally, the Gen 4 HD as the latest and currently best performing \ac{EBC} for \ac{SSA} is yet to be assessed in the same manner.


In this paper, the first extensive and purely \ac{EB} star catalogue analysis is developed and analysed using real-world data. This work captures the limitations of the \ac{EBC} in addition to the complex spatio-temporal features and characteristics of sources observed using the Gen 4 HD \ac{EBC} at different scanning rates, seeing conditions and weather conditions. The first automatic star mapper using only real-world \ac{EB} observations is developed in this paper to conduct this analysis. By analysing the \ac{EB} features of each event source and their relationship with the underlying photometric characteristics of the associated source, future exploration of \ac{EB} data processing systems and algorithms can be based on a more robust understanding of \ac{EB} target-measurement dynamics and \ac{EB} space imaging limitations. 

\section{Methodology and Data Collection}

\begin{figure}[]
    \centering
    \subfloat[]{\includegraphics[width=\textwidth]{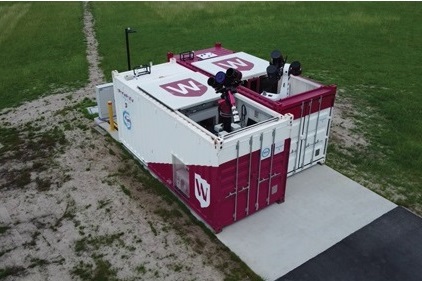}}%
    \qquad
    \subfloat[]{\includegraphics[width=\textwidth]{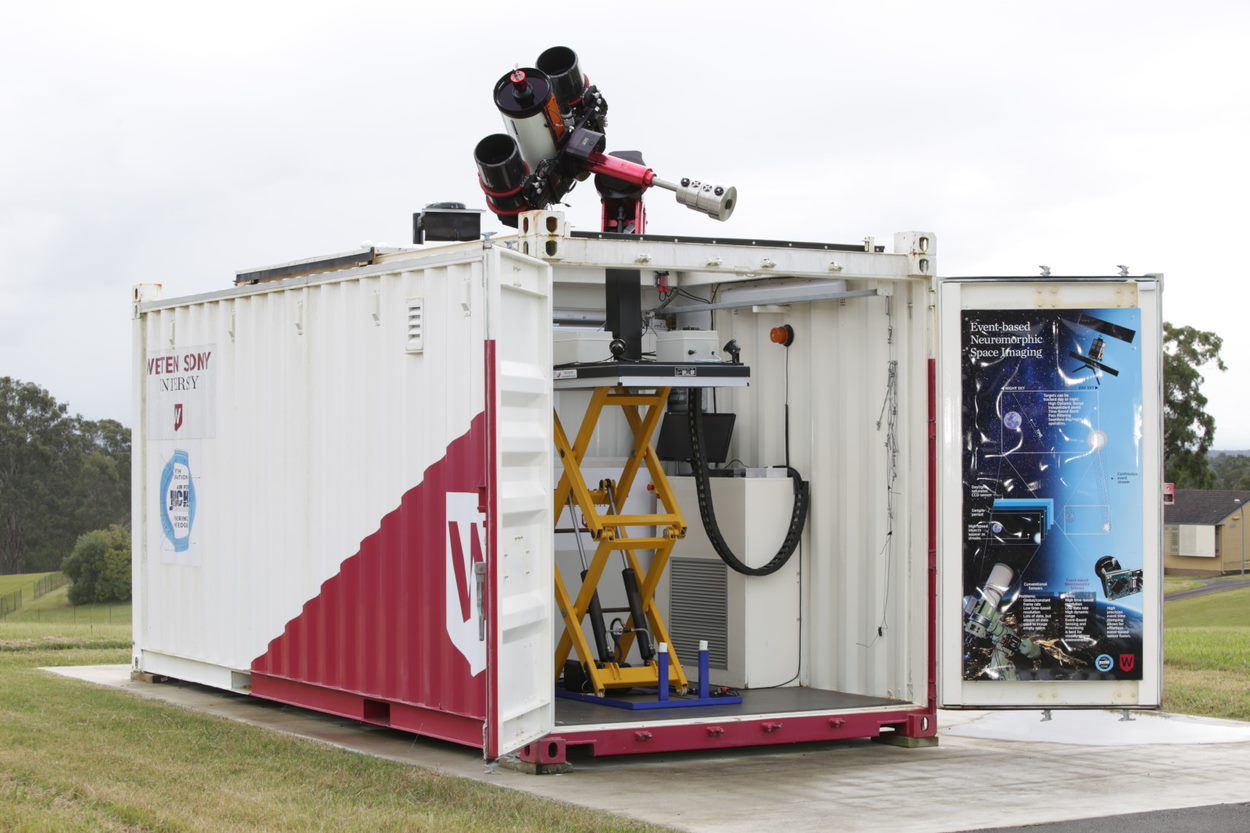}}%
    \caption{Astrosite-1 and Astrosite-2 (left and right respectively) at the Werrington North Campus (a) and the interior of the Astrosite-1 observing room (b).}
    \label{fig:astrosite}%
\end{figure}

This analysis is conducted on a dataset of real-world observations containing constant speed slews through a series of dense star fields using the Gen 4 HD in Astrosite-2. The Astrosite mobile observatories are built from standard 20~ft shipping containers, each with a control and telescope room. The mount and lifting mechanism is housed in the telescope compartment, where the lift raises the telescope mount out of the container via a retractable ceiling during observation, as in Figure \ref{fig:astrosite}. Operation of the Astrosite can be performed in the control room or remotely. As a mobile platform, the Astrosite can be rapidly moved to new observing sites to change the total effective coverage area, also known as \ac{FOR}. Since the telescope mount lifting mechanisms and the container itself are not sufficiently rigid or built into a structural foundation, the Astrosite imaging systems are more prone to movement and instability during observation than an observatory-based telescope. However, the high temporal resolution of the \ac{EBC} allows for compensation of this motion without the blurring effect that motion induces on conventional cameras. Furthermore, since the \ac{EBC} is a contrast-based detector, the additional mutual motion caused by vibration and instability can increase contrast in the scene.



Astrosite-1 was the first generation Astrosite observatory, which contains a Software Bisque Paramount ME II equatorial mount supported by a pneumatic scissor lift. Astrosite-1 is outfitted with an Officina Stellare RH200, which has a 200~mm aperture, 600~mm focal length and effective \ac{FOV} of $35.52\times20.05$~arcseconds with the Gen 4 HD \ac{EBC}. Astrosite-2 is the second-generation mobile observatory. The imaging system comprises a Planewave L600 Alt-Az mount on a sturdier pneumatic lift than Astrosite-1 and outfitted with the same telescope. The specifications of the mounts in each Astrosite are described in Table \ref{tab:chapter_2_astrosite_mounts}.

This variable speed \ac{EB} observation dataset\footnote{\textbf{The Event-Based Space Imaging Slew Speed Dataset is available from} \url{https://github.com/NicRalph213/ICNS_NORALPH_Event_Based-Space_Imaging-Speed_Dataset}} comprises 11 observations per field for a duration of 10 seconds each, with slew speeds varying from 0.5 degrees per second (or $1200\%$ sidereal rate), decreasing by $50\%$, to a minimum of 0.00048 degrees per second (or $11.5\%$ sidereal rate), where the sidereal rate is given as 0.00418 degrees per second.

\begin{figure*}
    \centering
    \includegraphics[width=\textwidth]{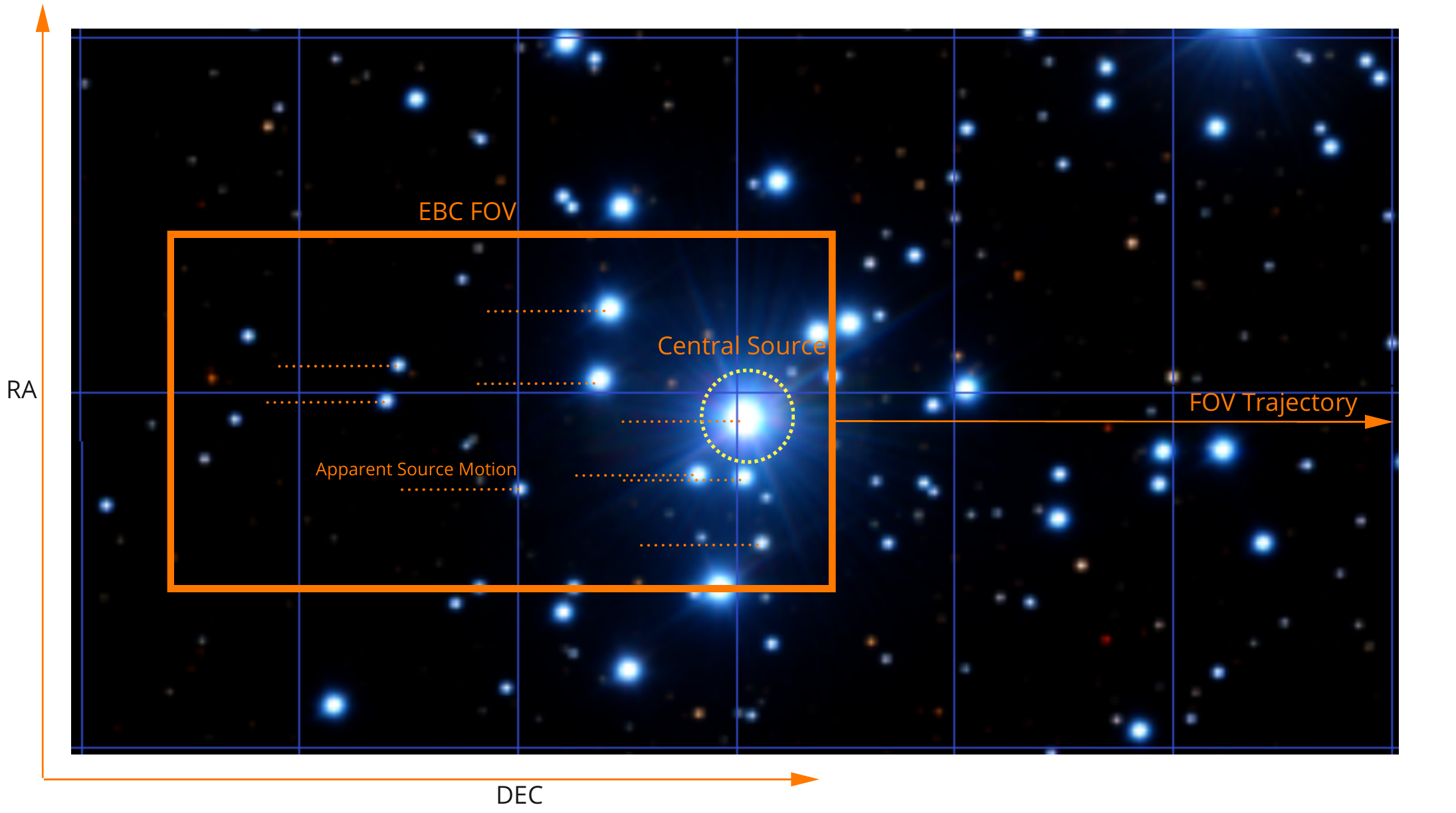}
    \caption{Demonstration of the speed survey observations with the apparent motion of the EBC FOV against a background star field (Jewel Box, NGC 4755) during a constant speed declination axis (DEC) slew. Star field generated using \texttt{Stellarium}.}
    \label{fig:star_survey_demo_stellarium}
\end{figure*}

A total of 4 fields were observed, each centred on a single bright source (approximately Mag $0$) among a dense star field with a variety of brightness to maximise source diversity, as in Figure \ref{fig:star_survey_demo_stellarium}, and summarised in Table \ref{tab:summary_speed_survey_fields}. At the start of the observation, the mount is slewed through the target field on the azimuth axis only (for smooth mount motion), reaching the centre at the approximate mid-point of the recording. In the case of the slowest slews, the recording begins with the source near the centre of the \ac{FOV} (given how much time it would take to move the mount at $11.5\%$ sidereal rate to transit the \ac{FOV}). 


Using a field velocity estimator, the event stream is motion-compensated for the scanning slew motion of the mount to collapse the observed event streaks into event sources in a global `star map' frame where sources resemble their apparent static appearance. The relative positions of the event sources can be used to perform an astrometric calibration of the field. Astrometric calibration serves as an estimate of the transformation between the pixel frame and the world frame. This transformation is used to identify sources in the star map using an established optical sky survey. Finally, real-world surveyed sources' photometric and physical characteristics are related to the spatio-temporal features and characteristics of sources in the star map. This analysis aimed to explore the following:

\begin{enumerate}
    \item Effect of slew rate on the spatial distribution, extent and event rate of an event source,
    \item accurate measurement methods of source centroid and spatio-temporal features of an event source,
    \item differing spatio-temporal features and dynamics of point sources and extended sources,
    \item dynamics introduced by varying wind and sky seeing conditions,
    \item limitations and trade-off between varying mount slew rates and the detectable contrast of sources.
\end{enumerate}
 

Three steps are required to conduct this analysis. Firstly, creating star maps from event streams. Secondly, detection and localisation of sources within the star map. Thirdly, astrometric calibration of the star map. Solving these challenges is critical in converting centroid measurements of \acp{RSO} in pixel coordinates to world coordinates to perform real-world \ac{SSA} tasks.

\section{Creating Star Maps from Event Data}

A star map is an image of sources or a table of source positions observed from a space imaging system. An event star map is an image frame compensated for the motion of the mount which contains the spatio-temporal features of sky sources. In such a frame, stars and astrophysical objects appear as fixed points. Producing a star map of event data enables analysis of the spatial features of these event sources and provides an avenue to perform astrometric image calibration on event data using established techniques. This image calibration facilitates the identification of astrophysical sources by extracting static source positions from the moving field and comparing their relative geometric arrangement to known surveyed star positions.

\begin{table*}[]
    
    \centering
        \begin{tabular}{|c|c|c|c|c|c|}
        \hline
        \textbf{Astrosite} & \textbf{Telescope Mount}                                                  & \textbf{Design} & \textbf{\begin{tabular}[c]{@{}c@{}}Maximum \\ Speed\end{tabular}} & \textbf{\begin{tabular}[c]{@{}c@{}}Pointing \\ Accuracy\end{tabular}} & \textbf{\begin{tabular}[c]{@{}c@{}}Mount Tracking \\ Accuracy\end{tabular}} \\ \hline
        Astrosite-1        & \begin{tabular}[c]{@{}c@{}}Software Bisque \\ Paramount MEII\end{tabular} & Equatorial      & 4 degrees/s                                                       & \textless 30 arcseconds                                               & $\sim$1 arcsecond                                                     \\ \hline
        Astrosite-2        & Planewave L600                                                            & Altazimuth      & 50 degrees/s                                                      & \textless{}10 arcseconds                                              & \textless{}0.3 arcsecond                                              \\ \hline
        \end{tabular}
        
    
    \caption{Telescope mount configuration in Astrosite-1 and Astrosite-2. }
    \label{tab:chapter_2_astrosite_mounts}

\end{table*}

\begin{table*}[]
    
    \centering
    
        \begin{tabular}{|c|c|c|c|c|c|c|c|}
        \hline
        Camera &
          \begin{tabular}[c]{@{}c@{}}Resolution\\ (pixels)\end{tabular} &
          \begin{tabular}[c]{@{}c@{}}Latency\\ ($\mu$s)\end{tabular} &
          \begin{tabular}[c]{@{}c@{}}Dynamic \\ Resolution\\ (dB)\end{tabular} &
          \begin{tabular}[c]{@{}c@{}}Minimum\\ Contrast \\ Sensitivity\\ (\%)\end{tabular} &
          \begin{tabular}[c]{@{}c@{}}Chip \\ Size\\ (mm$^2$)\end{tabular} &
          \begin{tabular}[c]{@{}c@{}}Pixel \\ Size\\ ($\mu$m$^2$)\end{tabular} &
          \begin{tabular}[c]{@{}c@{}}Fill \\ Factor\\ (\%)\end{tabular} \\ \hline
        DVS346 &
          $346\times260$ &
          20 &
          120 &
          14.2-22.5 &
          $8\times6$ &
          $18.5\times18.5$ &
          20 \\ \hline
        Gen 4 HD &
          $1280\times720$ &
          20-150 &
          $>124$ &
          11 &
          $6.2\times3.5$ &
          $4.86\times4.86$ &
          $>77$ \\ \hline
        \end{tabular}
        
    \caption{Specifications and comparison of an early generation Inivation DVS364 and next generation Prophesee Gen 4 HD ATIS EBC used in this paper \cite{gallego2019event}.}
    \label{tab:chapter_2_EBC_specs}

\end{table*}

\begin{table}[]
\centering
    \resizebox{\columnwidth}{!}{%
        \begin{tabular}{|c|c|c|c|}
        \hline
        \textbf{Field ID} & \textbf{Field Centre} &\textbf{ Central Source} & \begin{tabular}[c]{@{}c@{}}\textbf{Central Source}\\ \textbf{Magnitude}\end{tabular} \\ \hline
        0 & 191.930378, -59.688764 & \begin{tabular}[c]{@{}c@{}}Mimosa \\ HIP 62434\end{tabular} & 1.25 \\ \hline
        1 & 125.628542, -59.509483 & \begin{tabular}[c]{@{}c@{}}Avior \\ HIP 41037\end{tabular}  & 4.2  \\ \hline
        2 & 144.302803, 6.835782   & \begin{tabular}[c]{@{}c@{}}10 Leo\\ HIP 47205\end{tabular}  & 5.0  \\ \hline
        3 & 182.103152, -24.728782 & \begin{tabular}[c]{@{}c@{}}Alchiba\\ HIP 59199\end{tabular} & 4.0  \\ \hline
        \end{tabular}
    }
\caption{Summary of each field in publicly available speed survey dataset collected using Astrosite-1.} 

\label{tab:summary_speed_survey_fields}
\end{table}

Consider a star field observation collected using an \ac{EB} space imaging system. While the sensor is tracking sidereally and is stationary with respect to a stellar source, the pixel projection of these sources on the celestial sphere is static. In this case, event sources will appear as static and condensed blob-like features with contrast being produced only by atmospheric effects. While undergoing a non-sidereal translation, such as an angular slew, these stellar sources translate and appear with greater contrast due to the mutual motion of the mount. In this case, the angular slew motion of the mount is a rotation projected onto a local tangent plane on the celestial sphere, which causes sources to appear as streaks translating within the image frame and the spatio-temporal event stream. While tracking an \ac{RSO}, background sources appear as streaks. At the same time, the mount is centred on the moving target, which bears the appearance of a stationary source, as demonstrated in Figure \ref{fig:demo_2dvs3d_rso_observation_chapter2}. Two `field velocities' are present in this scenario, one for the background sources and the other for the tracked target. If compensated for the target's motion, the target would appear stationary, while the other sources as streaks. If compensated for the background sources, the target appears as a streak, and the background sources appear as static and condensed event sources. To accurately understand the position of the \ac{RSO} with respect to the background sources during mount tracking, a system for star map generation and background field velocity compensation must be designed.






Event star map images can be created by compensating for one of the field velocities of an \ac{EB} observation. By transforming or `warping' events by the inverse of the translation velocity, the event streaks are collapsed along their trajectory into features which resemble a spatial estimate of the underlying source's \ac{PSF}. These event sources now comprise an event star map. Calculating the motion-compensated position $e'_k$ of an event $e_k$ by warping on a star map is trivial, and can be simplified using a constant field velocity hypothesis $\theta_v = V = [v_x, v_y]^T$ to create a single star map frame. This is especially the case in the speed dataset, where the mount slews with a constant velocity. Consider the event stream $\mathcal{E}$, comprising a set of individual events, $e_k$ with pixel position on the image plane $x, y$ at time $t$ with contrast polarity $p$:

\begin{equation}
    e_k = (x, y, t, p)
\end{equation}

Using notation similar to \cite{gallego2018unifying}, the warped event $e'_k$, is a function of a velocity hypothesis $\theta_v$, (field velocity $V$ in this constrained star mapping case) at some reference event timestamp $t_k$, 

\begin{equation}
    e'_k = W(e_k, t,;\theta)
\end{equation}

In this constrained case on the star map,

\begin{equation}
  e'_{k} = 
      \begin{bmatrix}
      x'_{k}\\
      y'_{k}
      \end{bmatrix} 
      =
      \begin{bmatrix} 
      x_{k}\\
      y_{k}
      \end{bmatrix}
      - 
      \begin{bmatrix} 
      v_{x_{k}}\\
      v_{y_{k}}
      \end{bmatrix}
      (t_{k}-t_{0})
\end{equation}\label{eq:warping_equation}

The star map image frame of warped events $H$ can then be calculated using $N_e$ events from the event stream and using the known warping hypothesis:

\begin{equation}
    H(e_k;\theta) = \sum_{k=1}^{N_e}b_k\delta(e'_k)
\end{equation}

Where $b_k$ is the accumulation value of warped events if they are warped onto the same projected pixel region ($\delta$). In this experiment, both star maps with negative polarity events ($b_k = p_k$) and without negative polarity events ($b_k = 1$) are analysed.  


In the literature, event warping and motion compensation using the \ac{EBC} is well studied for conventional applications, with the field velocity estimation method as the distinguishing factor between each study. The contrast maximisation approach is most widely used, where point trajectories in the event stream are estimated by maximising an objective function $f(\theta_v)$, to produce a set of suitable warping parameters $\theta_v$. The most common approach to estimating the warping parameters is uses the pixel value variance of the warped image \cite{gallego2018unifying} and maximum accumulation of pixel values on the warped image \cite{stoffregen2019event}.



An detailed review of these objective functions is presented in \cite{gallego2019focus}. Various optimisation approaches have been explored to locate the optimal $\theta_v$, ranging from gradient ascent \cite{stoffregen2019event}, branch-and-bound \cite{liu2020globally}, \ac{RANSAC} optimisation \cite{wang2021visual}, and gradient descent using neural networks \cite{zhu2019unsupervised}.



Although generic event warping and motion compensation has a strong foundation in the \ac{EB} literature, few works have focused on space-based applications. In \cite{cohen2018approaches}, a star map is generated using basic motion compensation with Equation \ref{eq:warping_equation} on real-world space imaging data using a manual field velocity estimation. Algorithms for automatically calculating the field velocity to create star maps have been achieved using the contrast maximisation technique \cite{chin2019star}, in addition to motion compensation by estimating the field velocity using target tracking with Hough line detection \cite{bagchi2020event}. Both \cite{chin2019star} and \cite{bagchi2020event} used physically simulated data. In \cite{mcmahon2021commercial}, integrated event frames are used instead of star maps for analysis and astrometric calibration. Instead, conventional images are co-collected using a \ac{CMOS}, with the scanning mode observations being astrometrically calibrated by calibrating the conventional image and then correlating the calibrations with the event frame. These studies produce star maps capable of astrometric calibration. However, no study has been published using an automatic algorithm to estimate field velocity to generate star maps using real-world data. 



This paper solves the field velocity estimation problem from an alternate and simplified tracking perspective. Since all background sources in a field are static on the celestial sphere, they exhibit a constant apparent motion consistent with a constant slew motion. Observations in the dataset used in this paper have constant motion and one significantly bright source, which transits the \ac{FOV} at the middle of the recording. Using these constraints, the field velocity can be simply calculated as the highest likelihood single target measurement association hypothesis for the largest target using a basic \ac{GNN} \ac{STT} state estimator. This approach is similar to \cite{ralphreal}, where so-called 'leap-frog observing' was used to constrain the tracking task to a more simple \ac{STT} task. The global single target track hypothesis is given by the state $x_k$, of the clustered source in the source set $\mathcal{X}$, at time step $k$, with the largest extent $E$, in the \ac{FOV} is given by,  

\begin{equation}
    x_{k} = \mathcal{X}_k^{argmax(E)}
\end{equation}

The field velocity $\theta_v$, is given by the position state change of the largest source between time intervals, 

\begin{equation}
    \theta_v = \frac{x_k - x_{k-1}}{t_k - t_{k-1}} 
\end{equation}

With this single global hypothesis, the field velocity of observation segments can be estimated with reasonable accuracy by assuming the state velocity estimate of the largest target at each event time-step is constant for the observation duration. Abstracted in Algorithm 1, multiple frames are processed by the \ac{STT} within 3 seconds before and after the midpoint of the observations, each with an integration time listed in the integration time of the velocity estimation frame, which is accumulated at the velocity estimation frame interval. The frame integration times and intervals used by the \ac{STT} are given in Table \ref{tab: velocity frame and star map integration times}, where slower fields require longer frames for accurate tracking. The integration time for star maps is 3 seconds for all fields and slew speeds. The final field velocity is calculated as the mean target velocity of the largest target over all intervals. Outlier velocity estimates beyond three sigmas of the mean are filtered. Using the \ac{STT} estimated field velocity, and assuming this velocity is constant throughout the observation, a star map can be created by warping the event stream accordingly. Detecting the event sources in the star map for astrometric calibration and spatio-temporal feature analysis still requires a source-finding algorithm to accurately localise event sources within the star map.

\begin{figure}
    \centering
    \includegraphics[width=\textwidth]{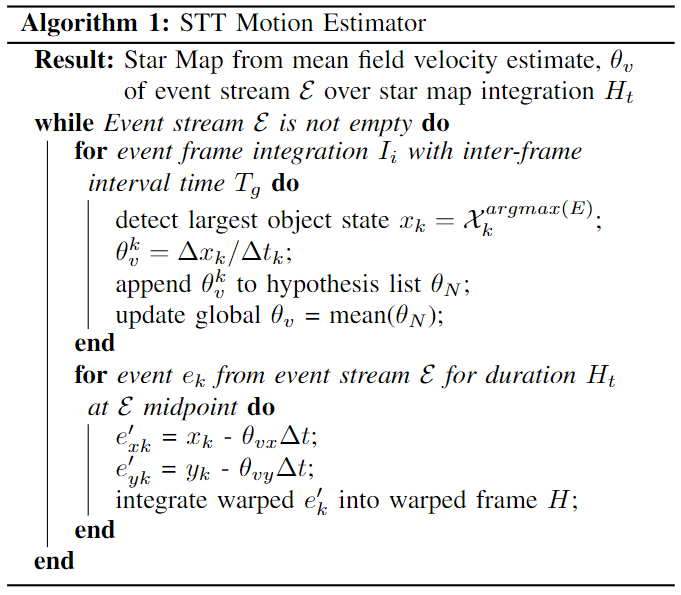}
    \caption{}\label{fig:stt_field_velocity_estimator}
\end{figure}



\begin{table}[]
\centering
    \resizebox{\columnwidth}{!}{%
        \begin{tabular}{|c|c|c|c|}
        \hline
        \textbf{\begin{tabular}[c]{@{}c@{}}Speed Range\\ (deg/sec)\end{tabular}} &
          \textbf{\begin{tabular}[c]{@{}c@{}}Velocity Estimation \\ Frame Integration \\ Time (s)\end{tabular}} &
          \textbf{\begin{tabular}[c]{@{}c@{}}Velocity Estimation\\ Inter-Frame \\ Interval Time (s)\end{tabular}} &
          \textbf{\begin{tabular}[c]{@{}c@{}}Star Map \\ Integration \\ Time (s)\end{tabular}} \\ \hline
        \textgreater{}0.07 & 0.05 & 0.05 & 3.00 \\ \hline
        \textless 0.07     & 0.05 & 0.25 & 3.00 \\ \hline
        \textless{}0.002   & 0.05 & 2    & 3.00 \\ \hline
        \end{tabular}  
    }
\caption{The integration times for the individual accumulated frames used by the STT to estimate the field velocity and to compose the star map. All fields and slew speeds use the same velocity and star map integration time. }
\label{tab: velocity frame and star map integration times}
\end{table}

\section{Localising and Analysing Event Sources in the Star Map}

Astronomical source finding is the process of locating and identifying sources as Gaussian \acp{PSF} on the image plane amongst background noise \cite{bertin1996sextractor}. Typically, source finding is performed using statistical techniques such as image sigma-clipping to isolate source \acp{PSF} amongst background noise. In the star mapping and calibration pipeline developed in this paper, sources are detected in a sigma-clip filtered star map using the density-based clustering algorithm \ac{DBSCAN} and pixel connectivity. Since the star mapping process produces an accumulated event frame, frame-based techniques similar to conventional astronomy can be performed. Clustering algorithms such as \ac{DBSCAN} can be used on event data to cluster spatial characteristics (as would be the case in conventional imaging). However, \ac{DBSCAN} can also cluster based on temporal characteristics using the associated time stamps of each compensated event in the frame. Performing higher-dimensional clustering using source temporal information improves source clustering by preventing spurious associations between events with high spatial similarity but low temporal similarity.       


The commonly used sigma-clipping method is implemented in this paper to filter regions in event star maps. Sigma-clipping is a traditional technique in astronomical signal processing, where values in a set with a target numerical characteristic are removed or `clipped' if the characteristic value lies outside of a standard deviations boundary from the global mean of the set (example in \cite{kleyna2004photometrically}). This process is iterated until a stopping criterion is met, such as an iteration limit, as detailed in Algorithm 2. In space imaging, sigma-clipping is especially useful in source finding, where sources are detected as statistically significant intensity peaks. Sigma-clipping is also used to remove pixels or elements with a value far too high (representing hot pixels, cosmic rays or other artefacts) or too low (background noise). In this project, sigma-clipping is used to mask accumulated pixel regions in the final star map to remove isolated events which likely represent background noise that could otherwise be included in values associated with a genuine source. 

\begin{figure}
    \centering
    \includegraphics[width=\textwidth]{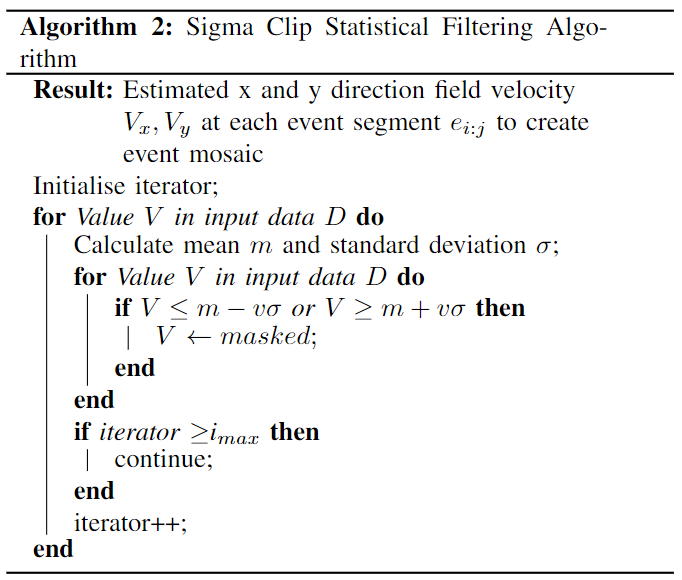}
    \label{alg:dbscan}
\end{figure}




\subsection{Source Finding Events using DBSCAN Event Clustering and Star Map Pixel Connectivity Analysis}\label{sect:source_finder}

The \ac{DBSCAN} algorithm \cite{ester1996density} is a powerful unsupervised clustering algorithm commonly used to associate points based on density. \ac{DBSCAN} is used here to associate events in the motion-compensated star map to detect event sources and noise in the full 3D $x, y, t$ spatio-temporal event space. \ac{DBSCAN} can cluster data with an arbitrary number of clusters, spatial extents, densities and shapes in the presence of noise \cite{schubert2017dbscan}. By using \ac{DBSCAN}, clustering can be performed without priors on the number of sources in the field, produce clusters robust to noisy events, and locate faint low-density source clusters. \ac{DBSCAN} requires only two parameters, the minimum number of points for a cluster to not be considered noise and a minimum association distance $\epsilon$. Using \ac{DBSCAN}, clusters are detected by selecting central core points, then detecting points within the $\epsilon$ distance, and assigning cluster or noise labels if the number of points is greater than the specified minimum number of points $P_{min}$. Heuristics for choosing \ac{DBSCAN} properties are relatively simple \cite{karami2014choosing}. $P_{min}$ should be set greater than the number of dimensions in the input data to prevent isolated points from being assigned a label. This parameter should be further increased from the default value of 4 while clustering higher dimensional or highly noisy data \cite{schubert2017dbscan}. It has also been suggested that a $P_{min}$ value twice the number of dimensions is ideal \cite{martins1999chromosomal}, which is the value used in these experiments. However, the $\epsilon$ distance threshold is more difficult to select and requires knowledge about the data to determine how close points would ideally reside. A costly adjacency matrix is also required for a high $\epsilon$. In this star mapping pipeline, an $\epsilon$ value of 3 is selected, given source events should have a mutual distance of 3 pixels at least, which is 4.76 arcseconds using the Officina \ac{RH} telescope setup in Astrosite-1.

\ac{DBSCAN} has been shown to perform well in \ac{EB} vision tasks with event frames \cite{hinz2017online} \cite{chen2018neuromorphic}. Furthermore, it can process data similar to event streams with an arbitrary number of low density and high noise clusters in the high dimensional spatio-temporal space. For practical reasons, however, \ac{DBSCAN} cannot be used to cluster events from extended sources. This limitation is due to the high event rate caused by the brightness and sizeable spatial extent of extended sources, which results in excessive memory usage due to \acp{DBSCAN} $O(n^3)$ complexity, largely comprising the adjacency matrix calculation to perform the $RangeQuery$. Extended event sources with high event rates are associated based instead on pixel connectivity to overcome this limitation.



Pixel connectivity is a fundamental conventional computer vision analysis technique for analysing the relationship between discrete pixels within a neighbourhood \cite{soille2008constrained}. Pixel connectivity is used three times in this pipeline. 
Firstly, to detect centroids of large extended sources as large streaks in the uncompensated star map for velocity estimation (if present). Secondly, as a less computationally intensive way to cluster the large volume of events that accompany extended sources compared to \ac{DBSCAN}. Lastly, as a final step in the source finder, define the spatial extent of sources and calculate the spatio-temporal characteristics of event sources such as area, extent, and weighted centroid. Source weighted centroids are used later as precise source positions for astrometric calibration. This pixel connectivity analysis is conducted on a sigma-clipped binary image of either streaks or the motion-compensated star map (depending on where within the pipeline it is used). An 8-connected Moore neighbourhood is used here, which implies events connected as immediate neighbours (including diagonally), belong to the same neighbourhood and originate from the same source. 



\section{Astrometric Calibration of Event Star Maps to Project \ac{EBC} Measurements to World Coordinate Frames}

In conventional optical astronomy and \ac{SSA}, data is collected with a frame-based image sensor such as a \ac{CCD} by exposing the sensor to the sky position of a target and collecting light from the target source and the surrounding background sources. Exposure times vary based on the task and the target's brightness, with faint targets requiring up to 10 seconds to collect sufficient light, while brighter targets may only require 0.5 seconds. Using the reported field centre position and known kinematics of the mount, the known pixel scale based on the optical configuration, and the physical orientation of the sensor in the telescope, the target's position in the collected image can be determined in the world frame with reasonable accuracy. The localisation accuracy can be further improved by comparing the relative position of the background sources surrounding the target to an established astrophysical survey. This comparison is performed using astrometric calibration \cite{pier2003astrometric}, which is an image calibration technique used in astronomy to match a set of observed sources to a star catalogue to determine the world position of the centre, angular orientation and pixel scale of an image. In the case of \ac{RSO} observations, the specific world coordinate frame could be expressed as a \ac{WCS} or \ac{ICRS} coordinate frame, with the observer position in a geographic coordinate frame.

The ideal approach to converting event data from pixel coordinates to a world reference frame is the same as conventional space imaging. The dilemma, however, lies in the high temporal resolution of the \ac{EBC}. Events can be reported approximately every 20~$\mu$s, which is significantly faster than the 500~ms latency of the mount's position reporting. This temporal resolution is typical of most telescope mounts since mount position estimates on a shorter timescale than this are usually not required with the relatively long duration of \ac{CCD} exposures. With the high temporal resolution of the \ac{EBC}, any high-frequency changes in the mount position due to wind (which the mount is prone to given its mounting on a lifting platform), mount mechanical dynamics or other factors which move the optics that are detectable by the \ac{EBC}. This motion cannot be accounted for without more continuous mount position reporting.

Furthermore, since Astrosite-1 is a mobile observatory with mounts supported by lifting mechanisms, the quality of the pointing models is poorer than more stable and fixed conventional observatories. This inherent instability further increases the need for accurate calibration to overcome inevitable pointing offsets. While the telescope mount provides positions for the geometric field centre, it already comes from an astrometric calibration calculated using a conventional camera while calibrating the mount. Termed \texttt{t-point} modelling, this calculation determines the forward and inverse kinematics of the mount with reasonable accuracy. However, a rotation is still present due to the physical rotation of the sensor in the telescope and a translation offset due to calibration errors and imperfections in the telescope bore sighting.

 As a result of these practical challenges, astrometric calibration must be used to determine the world projection of events to produce high-precision measurements. Often performed using `plate solving' in conventional astronomy \cite{lang2010astrometry}, this practice is vital in converting observation data from the pixel frame into meaningful measurements in a world reference frame.

\begin{figure*}
    \centering
    \includegraphics[width=\textwidth]{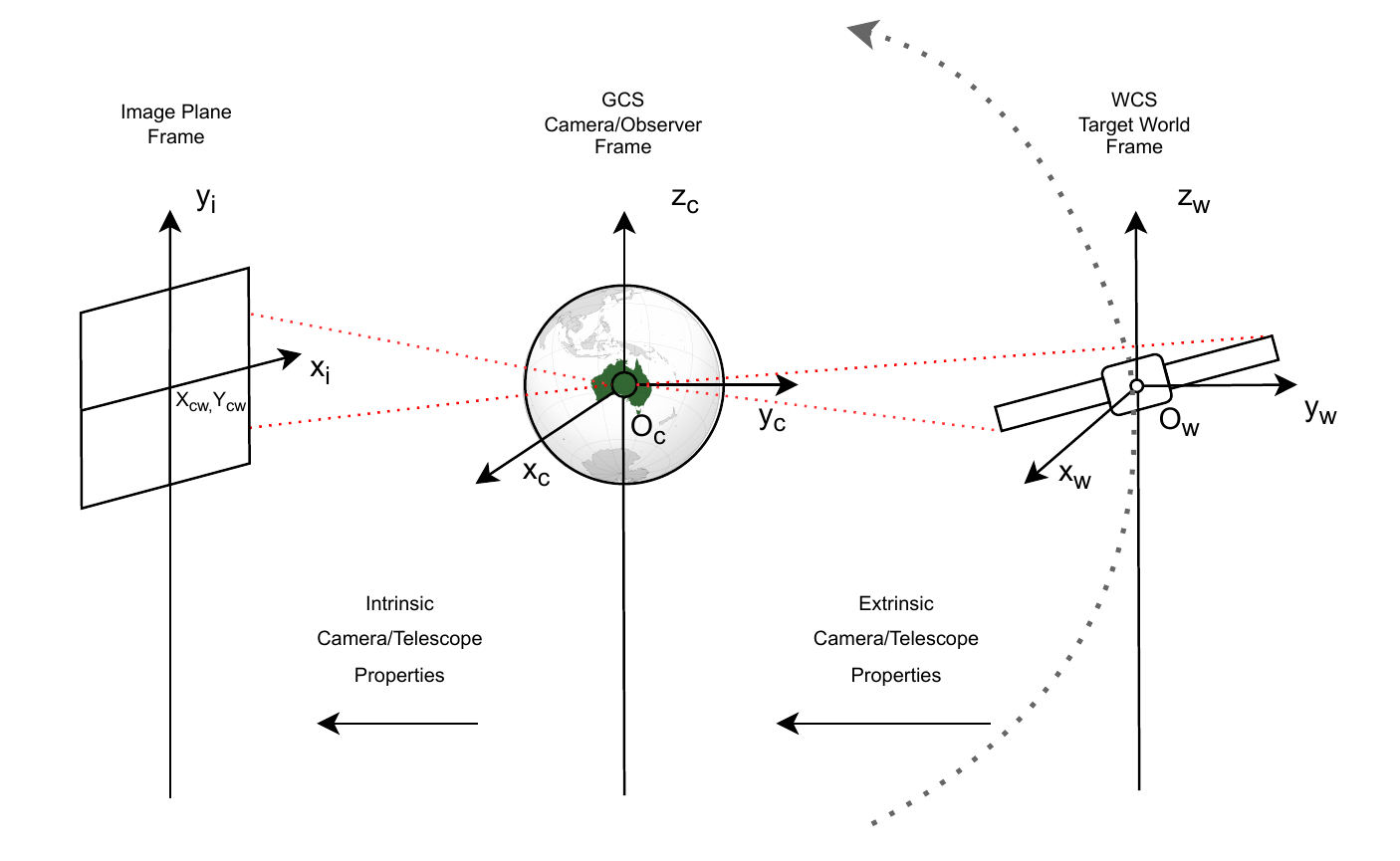}
    \caption{Demonstration of the transformation between an RSO in the WCS frame to a GCS camera coordinate frame using the extrinsic properties of the camera/telescope, and the transformation between the pixel coordinate frame on the image plane using the intrinsic parameters of the camera.}
    \label{fig:ssa_calibration_demo}
\end{figure*}

Astrometric calibration is a well-studied image calibration technique for understanding the transformation between a world coordinate frame to a camera coordinate frame. This technique uses the extrinsic properties of the camera and the transformation between the pixel coordinate frame on the image plane using the camera's intrinsic parameters, as illustrated in Figure \ref{fig:ssa_calibration_demo}. In optical and \ac{EB} astronomy, image calibration and homography is used to transform between frames $X$ to $X'$ using a standard homography matrix $H_{Rt}$ \cite{dubrofsky2009homography}:

\begin{equation}
   X' =  H_{Rt}X    
\end{equation}

The vector representation of a generic frame transformation using an assumed distortion-free pin-hole camera is given by this homography matrix as a homogeneous transformation matrix $H_Rt$ \cite{seedahmed2006direct}. This transform projects the image frame $X_i$, onto the world frame $X_w$, which comprises rotation $R_{3\times3}$ and translation $d_{3\times1}$ transformations, 

\begin{equation}
    \begin{bmatrix} X_i\\Y_i\\Z_i\\1 \end{bmatrix} = \begin{bmatrix} R_{xx} & R_{xy} & R_{xz} & d_x\\R_{yx} & R_{yy} & R_{yz} 
 & d_y\\R_{zx} & R_{zy} & R_{zz} & d_z\\f_x & f_y & f_z & \omega \end{bmatrix} \begin{bmatrix} X_w\\Y_w\\Z_w\\1 \end{bmatrix}
\end{equation}

In this space imaging scenario, the optics focus on infinity, and all sources are projected onto the celestial sphere. Consequently, a simplified approach is used for this scenario as a calibration between two 2D planes. Since the \ac{FOV} of the \ac{EBC} even in the relatively wide field \ac{RH} telescopes is no larger than 0.4 degrees, the observable effect of the celestial frame is minimal as the curvature of the frame is essentially zero. In this case, the $z$ axis can disregarded \cite{tsai1987versatile}. Further, lens distortion is not considered since the Astrosite uses reflective telescopes and no lenses are present in the optical path. The typical calibration transform for this idealised pinhole camera in 2D using `camera matrix' $P$ is given as, 

\begin{equation}
    \begin{bmatrix} X_i \\ Y_i \end{bmatrix} = P\begin{bmatrix} X_w \\ Y_w \end{bmatrix}
\end{equation}

Where the camera matrix $P$ describes the intrinsic transformation between the image plane $X_i$ and the camera frame $X_c$, and the extrinsic properties of the camera to transform between the camera frame and the world frame $X_w$. 

\begin{equation}
    P = \begin{bmatrix} f_x & 0 & c_x \\0 & f_y & c_y \end{bmatrix} 
    \begin{bmatrix} R_{1,1} & R_{1,2} \\R_{2,1} & R_{2,2} \\d_1 & d_2 \end{bmatrix} 
\end{equation}

 Using a plate-solving algorithm, these transformation properties can be estimated. The complete $X_i \to X_w$ transform can be expressed in simpler terms as a rotation matrix $R_{2\times2}$, pixel scaling factor $\omega$, and translation of the mount centre position $d_{1\times2}$. The mount centre translation is an important factor that describes the offset of the geometric image centre in pixel space to the telescope/camera frame origin, $X_w$. This transformation is given as, 

\begin{equation}
        \begin{bmatrix} X_w\\Y_w\end{bmatrix} =  \begin{bmatrix}\omega_x\\\omega_y\end{bmatrix}\left( \begin{bmatrix} R_{xx} & R_{xy}\\R_{yx} & R_{yy} \end{bmatrix} \begin{bmatrix} X_i\\Y_i \end{bmatrix} + \begin{bmatrix} d_x\\d_y\end{bmatrix}\right)+\begin{bmatrix} X_{cw}\\Y_{cw}\end{bmatrix}
\end{equation}

If a star map $H$, can be produced from the event stream, the calibration properties of the field and the estimated centre can be solved. Then the calibration properties can be applied to the pixels of each frame to project all events onto the world coordinate frame as meaningful measurements for \ac{SSA} tasks. The projected world position of the camera frame origin can be calculated by aligning the image plane frame to the camera coordinate frame through rotation and then translating the origin of the image plane frame to the camera coordinate frame origin. Finally, the new image plane frame is scaled by the pixel projection scale and translated to centre on the camera's origin in the world frame $X_{cw}$ (provided by the plate solver). The pixel scale using the square \ac{EBC} pixels is identical in both directions, $\omega = [\omega_x, \omega_x]^T = [\omega_y, \omega_y]^T$. Using the solved calibration parameters and a field centre from the plate solver $X_{cw}$, pixels $[X_i, Y_i] \in [e_{xk},e_{yk}]$ as events or points on a motion-compensated event star map frame $H$, can now be projected into the world frame, as in Figure \ref{fig:ssa_calibration_demo}.

If a star map is excessively large (>$10,000$ square pixels before source extraction), motion-compensated star map frames from the whole event stream can be continually created and calibrated. In this paper, this technique is coined as \ac{MPMI} calibration, which is the continuous calibration of the event stream using multiple solved star maps as a mosaic. Although computationally intensive, this approach provides low latency field solutions, enabling variable speed analysis and improving position reporting robustness to wind and other short interval effects that may move the optics and mount between the mount position reporting intervals. This calibration approach is only possible with sufficient detectable background stars to correctly estimate the field velocity for plate solving. Scenarios where the \ac{FOV} has few detectable sources are typically encountered when observing in poor seeing conditions or during high-speed tracking slews or where long intervals are not possible. These conditions limit the sensitivity of the \ac{EBC} and may result in insufficient detected sources for plate solving. In such cases, an alternate calibration technique is required. 
 
In this case, where a star map cannot be produced or solved, but the calibration properties are known from a previous successful calibration in more favourable conditions, a previous solution may be used. Any event within the input event stream can be projected onto the world frame using such a calibration solution. This transformation does, however, require the reported position of the mount as the assumed projection of the field centre in the pixel frame $X_{ci}$ in the world frame $X_{cw}$, and the field centre estimate $[\bar{X}_{cw}, \bar{X}_{cw}]^T$ to be valid. Additionally, previously calibrated estimates of the $R$, $d$ and $\omega$ can be used in this solution.

 Termed in this paper as \ac{SPSI} calibration, plate solving may be performed once per night on a single point under favourable conditions with sufficient sources and using an observing technique that maximises the detection limit, such as a slow tracking slew. With a sound pointing model, this blind approach to calibration \cite{wijnholds2016blind} assumes that the translation offset of the field centre and orientation is constant across the sky (except for mount motion effects such as wind which would require constant target tracking, as discussed later). Although \ac{SPSI} calibration is computationally cheaper than \ac{MPMI} calibration, this approach relies on using the most current field centre position reported from the mount. Since the temporal resolution of the \ac{EBC} is so high, the mount positions must be interpolated so that event positions in the pixel frame can be transformed to the world frame coordinates at the event time stamp. Interpolation is naturally less accurate and disregards most high-frequency and unexpected changes to the mount position. Corrections may also be required in calibration to account for atmospheric effects and mount motion. 





\subsection{Solving for Calibration Parameters of an Event Star Map using Astrometrynet}

In this paper, astrometric calibration is performed using the \texttt{Astrometrynet} \cite{lang2010astrometry} plate solver. \texttt{Astrometrynet} is an accurate and widely used package which can plate solve a sky image or tabulated set of source locations. \texttt{Astrometrynet} plate solves using four steps:

\begin{enumerate}
    \item Input or calculation of sub-pixel accurate source positions of stars within the input field by identifying statistically significant peaks in local image intensity,
    \item Calculation of a geometric hash for detected subsets of four stars or `quads',  
    \item Search a large pre-computed index of known stars for hash codes which match the quads to determine a hypothesised alignment between the quad in the query input image and a quad in an index,
    \item Verification as a Bayesian decision problem to accurately decide if the hypothesised alignment is correct and therefore determine a hypothesised location, scale and orientation of the input image on the sky. 
\end{enumerate}



\texttt{Astrometrynet} does not require priors, but information about the field can improve the solving time. Astrometric calibration using a local installation of \texttt{Astrometrynet} can be performed within seconds when provided with the known pixel scale of the \ac{EBC} (derived based on the optical configuration of the telescope), a rough pointing centre of the star map (obtained from the reported mount position) and a search radius scaled to an approximation of the observation area. A minimum of 4 sources are required by \texttt{Astrometrynet} for calibration if the approximate position and pixel scale are provided. The main output is stored in the \texttt{.fits} file contains the calibrated input image and a human-readable header with a summary of the calibration results. These results include the solutions required to perform the projection from the pixel frame to the world frame. These values are the pixel-\ac{WCS} transformation matrix, the pixel scale in the \ac{WCS}, the `CRVAL1/2' RA/DEC centre coordinates, the `CTYPE1/2' projection type, and the `CRPIX1/2' pixel projection and rotation reference point.

\subsection{GAIA DR2 External and Internal Source Cross-Matching}

To identify detected sources in the generated star maps as real-world astrophysical objects, this star mapping pipeline cross-matches the calibrated \ac{WCS} positions with the GAIA Data Release 2 (DR2) optical astronomy catalogue (the current largest optical survey integrated into \texttt{astrometrynet} \cite{gaia2018gaia}). The Gaia DR2 catalogue contains several source characteristics, namely the source identification and optical magnitude in multiple bands. Additionally, calibration is conducted with the `5200 Heavy' index, which contains sources from the Tycho-2 and Gaia-DR2 catalogue. This index is ideal for fields that are more narrow than 1 degree and contains additional Gaia-DR2 information such as magnitude and catalogue ID.

For each field, the extracted source characteristics are collated and internally cross-matched. Fields containing too few sources or that were observed during high-speed slews may not contain sufficient sources for a successful calibration. Since all observations of a given field are conducted with a slew from the same direction and angle, sources in the unsolved slews can be cross-matched to the solved slews using the common geometric arrangement of each source with respect to the central bright extended source in each observation. For all fields, the horizontal $x$ and vertical $y$ distances between each source and the largest source in the field are calculated. These distance measures are then used to associate sources within the same field. The most likely candidate match will have the closest mutual distance similarity. Inspired by the speed at which sources are matched in \texttt{Astrometrynet}, these matches are located using the nearest neighbour criteria on a KDtree binary tree. This technique is efficient and scales well with the $~500$ total source detections in each field, with an $O(n)$ complexity for storage operations and $O(log(n))$ to $O(n)$ for searching operations. Optimised approaches such as these are vital for future \ac{SSA} systems, which are likely to produce significantly more data and have a more significant design focus on low latency reporting. 

Potentially spurious sources detected in unsolved slews can be eliminated if no successful cross match is found (within a 5-pixel threshold in the $x$ and $y$ directions). `Internally' cross-matching using this method improves the overall robustness of the pipeline by providing every observation for a given field with an astrometric solution when at least one observation has been solved. However, spurious detections in solved fields themselves may not be de-correlated.  


\section{Overall Star Mapping, Source Analysis and Calibration System}\label{sect:star_mapper_overview}

The full star mapping algorithm and calibration pipeline developed in this paper is described below in Figure \ref{fig:chapter_3_calibration_pipeline}. The pipeline can be summarised as:

\begin{enumerate}
    \item Event star map generation using \ac{GNN} \ac{STT} velocity estimation,
    
    \item Noise filtering using sigma-clipping, 
    
    \item Extended source detection and masking using pixel connectivity,
    
    \item Point source detection using \ac{DBSCAN} event clustering,
    
    \item Spatio-temporal analysis of detected extended and point sources using pixel connectivity,
    
    \item Astrometric calibration of the field using an \texttt{Astrometrynet} plate solve of the source positions to project sources in the image frame to the world frame,
    
    \item Analysis of the relationship between spatio-temporal characteristics of the detected event sources and the photometric characteristics of the underlying astrophysical objects.
    
\end{enumerate}

\section{Results}\label{sect:measuring_sky_results}

\begin{figure*}[]
    \centering
    \includegraphics[width=0.75\textwidth]{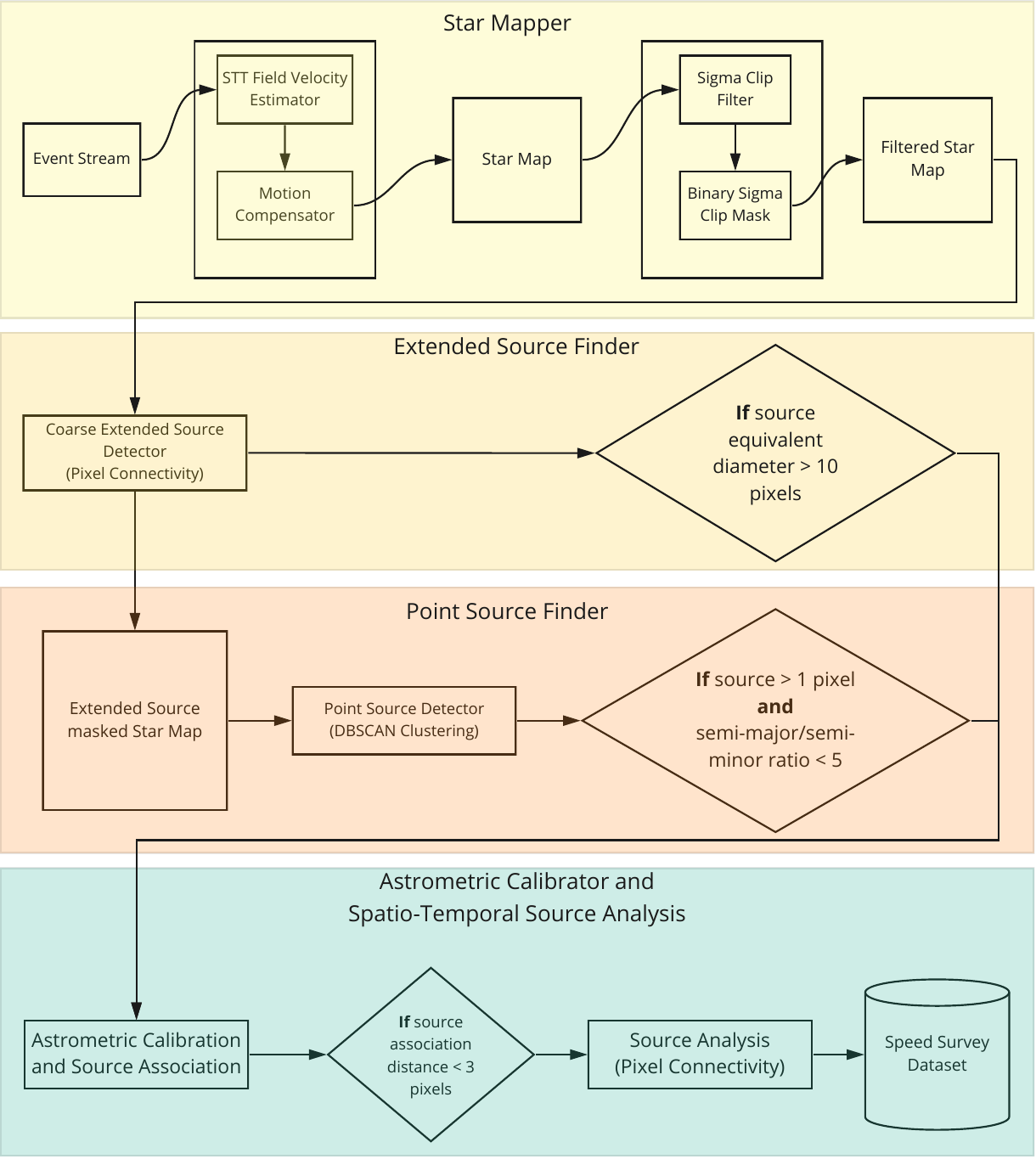}
    \caption{Overview of the star mapping, astrometric calibration and spatio-temporal event source analysis pipeline developed in this paper.}
    \label{fig:chapter_3_calibration_pipeline}
\end{figure*}

\renewcommand{\cleardoublepage}{}
\renewcommand{\clearpage}{}

\subsection{Star Mapper Performance}

\begin{figure}[]
    \centering
    \subfloat[]{\includegraphics[width=\textwidth]{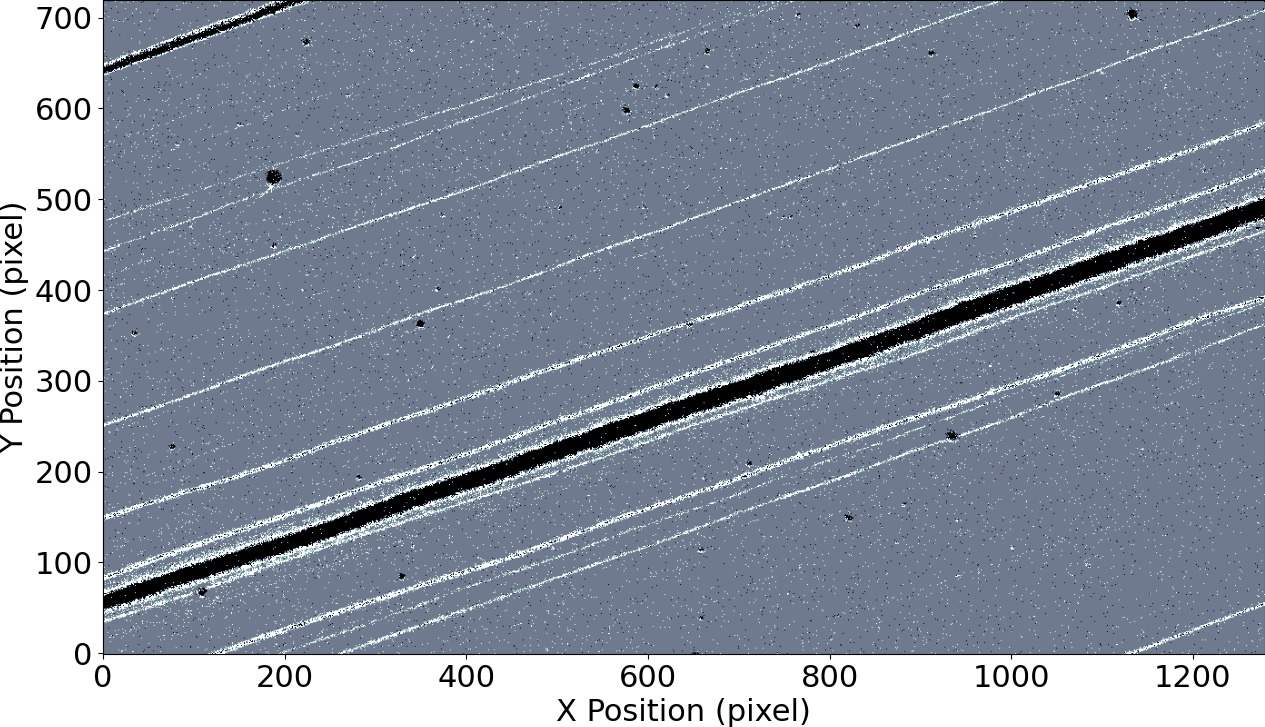}}%
    \qquad
    \subfloat[]{\includegraphics[width=\textwidth]{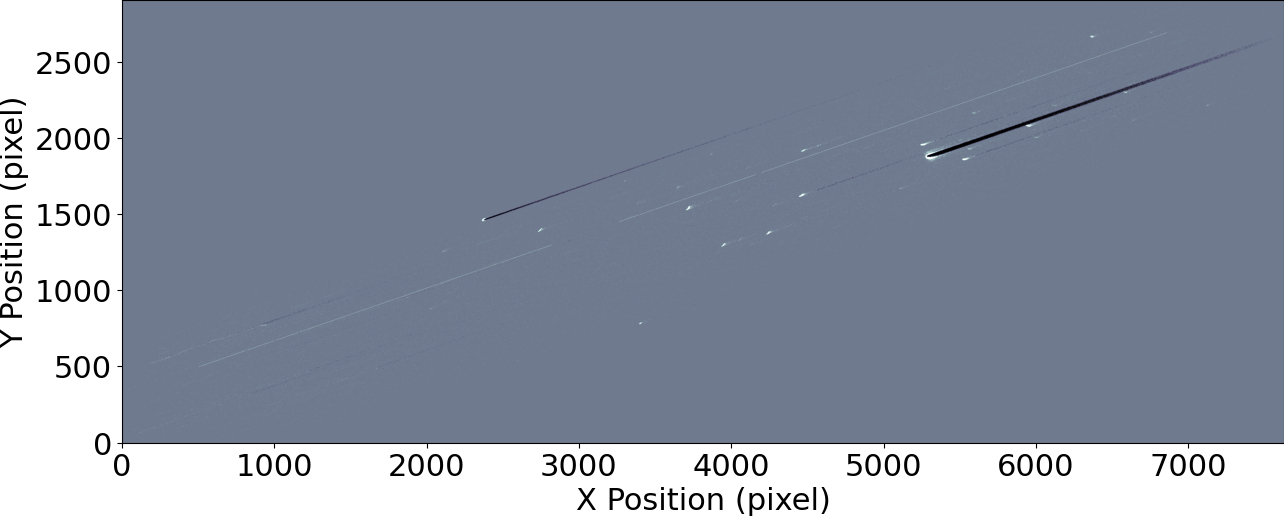}}%
    \caption{Comparison between the original 0.5-second event integration of Field 0 at the highest speed of 0.5 degrees per second (a), and the successfully motion compensated star map (b), where off-events are shown in black and on events in white.}%
    \label{fig:fast_accumulated_and_compensated}%
\end{figure}

\begin{figure}[]
    \centering
    \subfloat[]{\includegraphics[width=\textwidth]{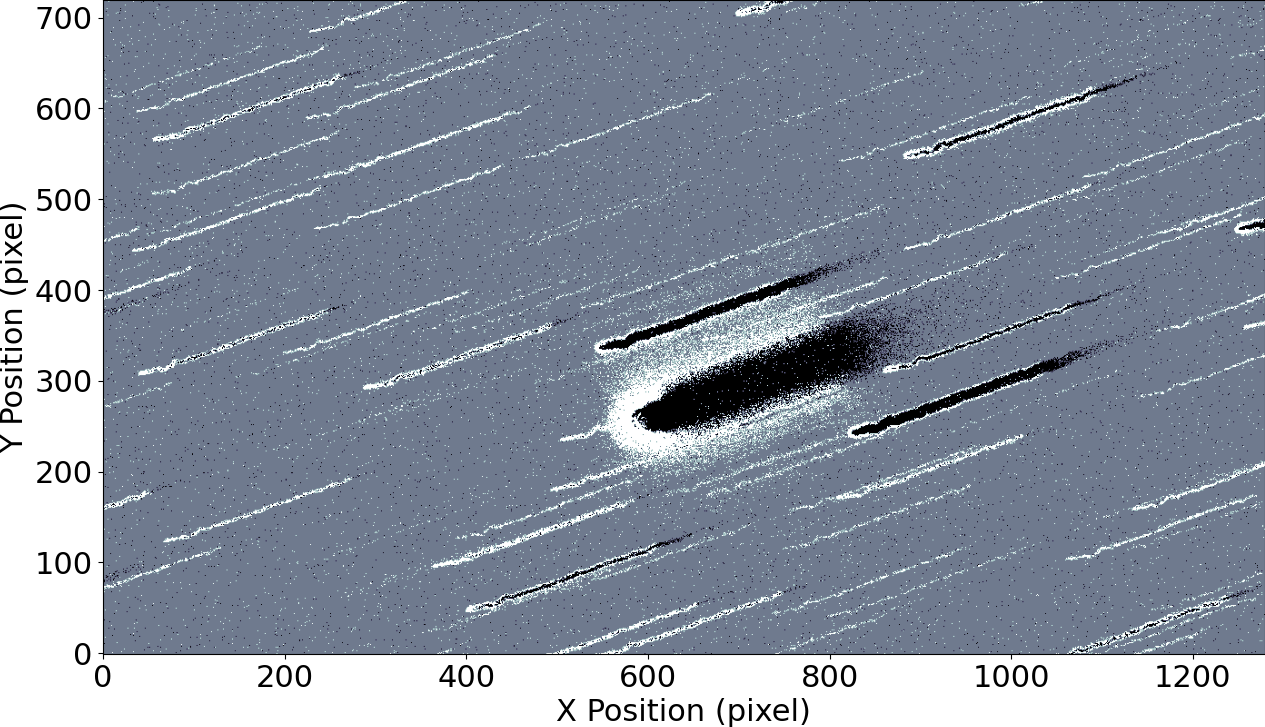}}%
    \qquad
    \subfloat[]{\includegraphics[width=\textwidth]{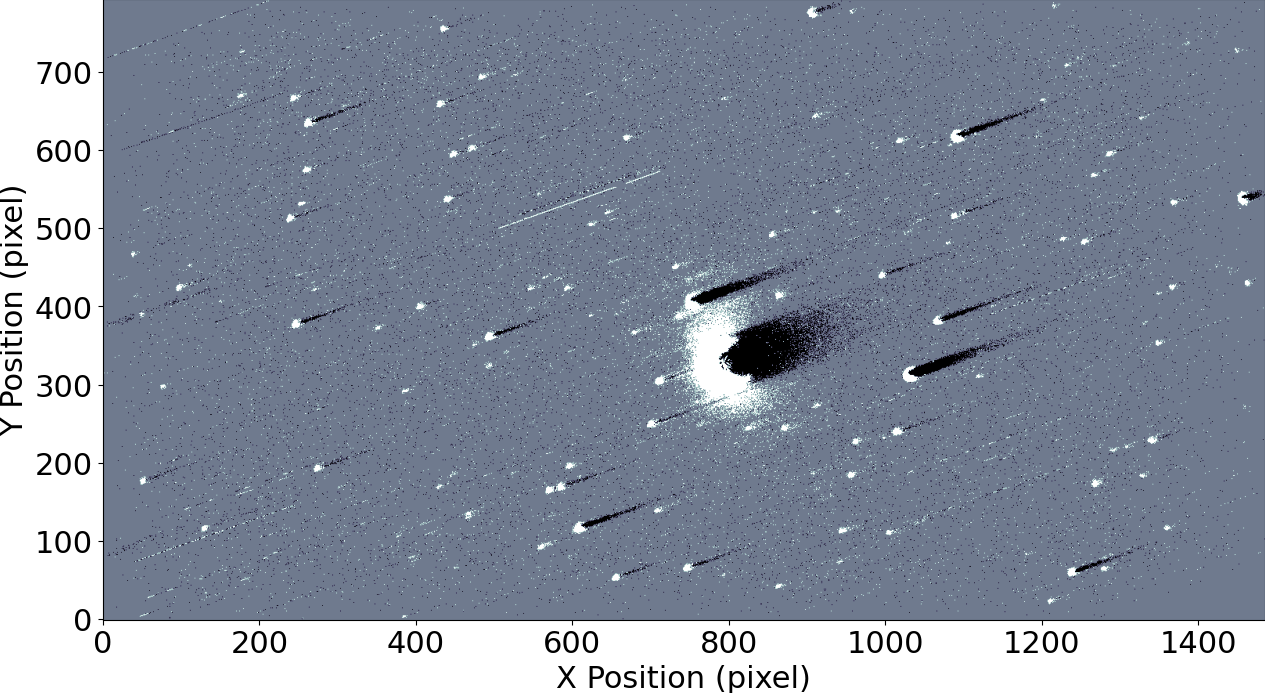}}%
    \caption{Comparison between the 1.5-second event integration of Field 0 at the mid-range speed of 0.015 degrees per second (a), and the correctly motion compensated star map (b)}%
    \label{fig:medium_accumulated_and_compensated}%
\end{figure}

\begin{figure}[]
    \centering
    \subfloat[]{\includegraphics[width=\textwidth]{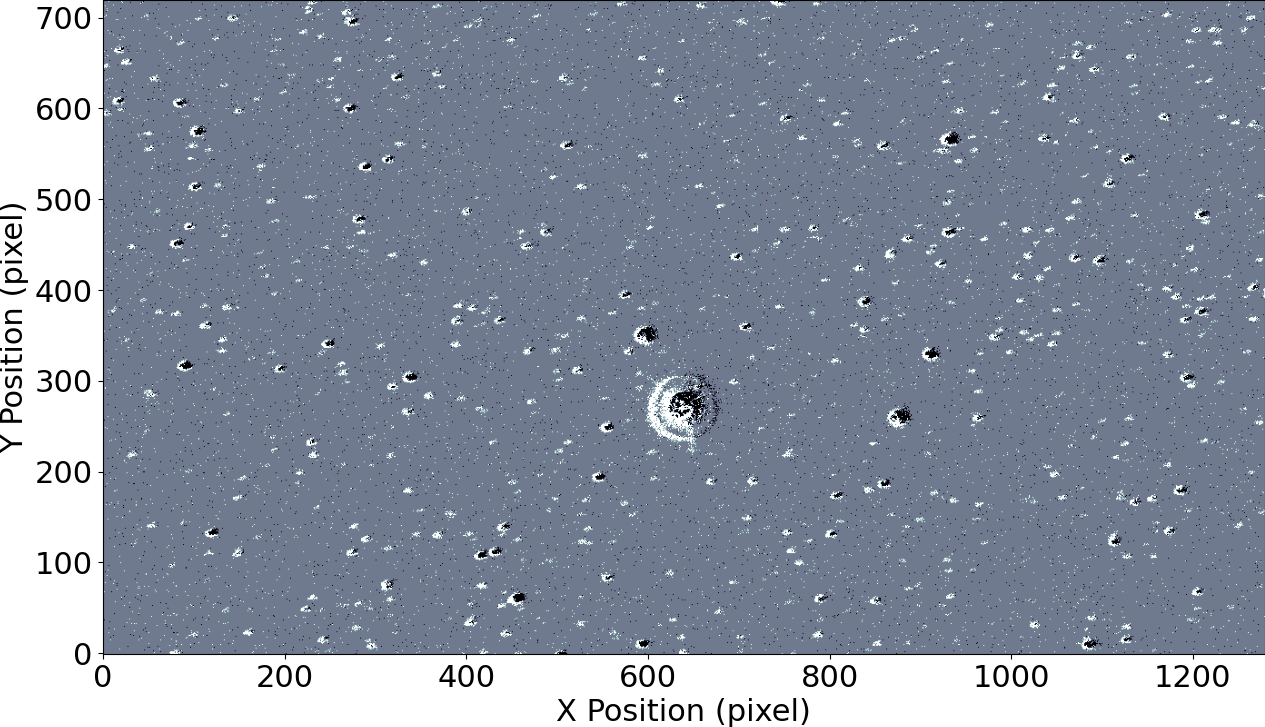}}%
    \qquad
    \subfloat[]{\includegraphics[width=\textwidth]{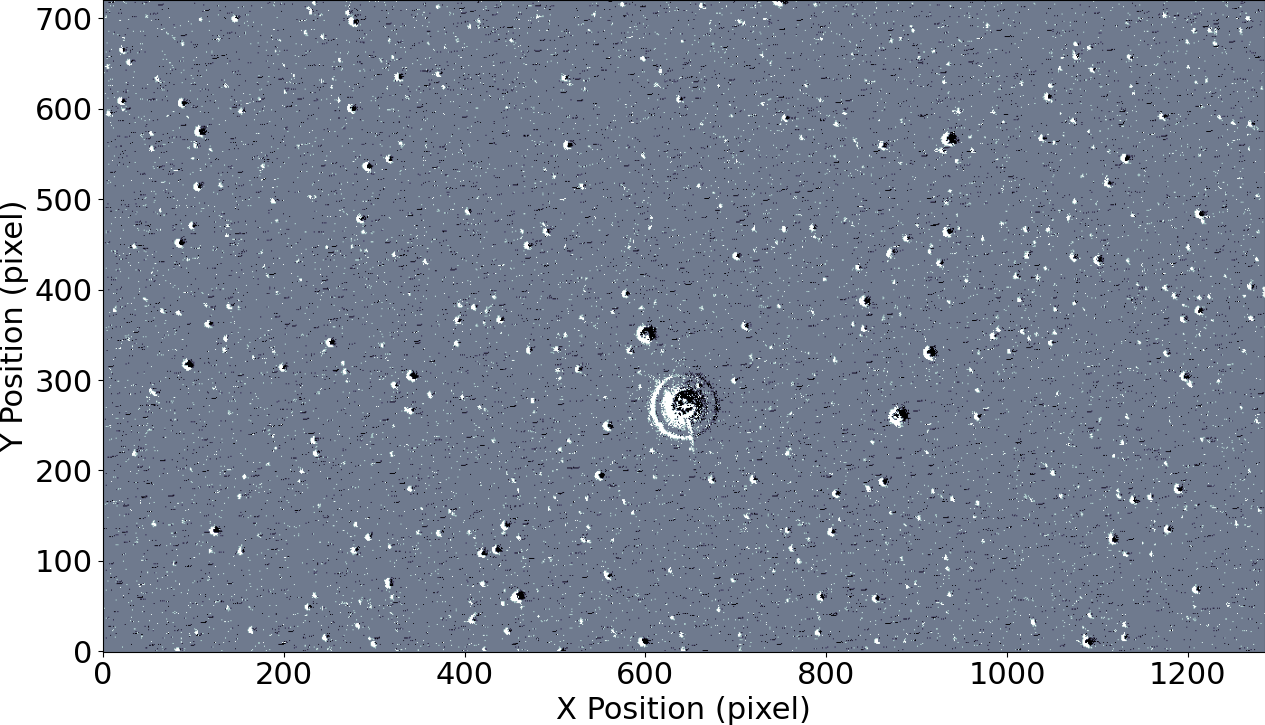}}%
    \caption{Comparison between the 3-second event integration of Field 0 at the lowest speed of 0.000488 degrees per second (a) and the correctly motion compensated star map (b), where the sharp features of the Mimosa Airy disk are now visible.}%
    \label{fig:slow_accumulated_and_compensated}%
\end{figure}

The star mapping algorithm developed in this paper is shown to successfully warp observations collected across four different star fields with nine different slew speeds to create star maps containing resolved event sources with sharp spatial features. In Figures \ref{fig:fast_accumulated_and_compensated}-\ref{fig:slow_accumulated_and_compensated}, the success of the \ac{STT} velocity estimation technique is shown by compensating for three field speeds in the Mimosa field, as the densest star field with the brightest central source and greatest diversity of source brightness. The accumulated events streaks over the integration period are warped onto their trajectory to produce a star field with compact event sources. These star maps contain a variety of sources with diverse magnitudes. Bright extended sources are immediately apparent in the high-speed observations, while faint sources become gradually more visible at the mid-range speed and low speed. In low-speed observations, sources appear to have few wake events and appear with a smaller spatial extent. Source brightness is apparent in these images, as brighter targets produce significantly more events and are larger than faint targets. 

The features of the bright extended sources evolve over each observation and slew speed to gradually resemble the Airy disk. In Figure \ref{fig:slow_accumulated_and_compensated}, the Airy disk is visible around Mimosa, in addition to small diffraction spikes, as expected from a nearly diffraction-limited imaging system (atmospheric effects are naturally still present), and the physical construction of the \ac{RH} Officina in Astrosite-1. In the slow-speed recordings, the on and off-events form two symmetrical halves of the Airy disk. In high-speed recordings, however, the Airy disk features are more blurred, and the off-events produce a long trail of wake events.

Streaks still present in the warped frames are stationary hot pixels, which are now stretched when compensating for the mount slew motion. In the fast slew accumulated frame, Figure \ref{fig:fast_accumulated_and_compensated} (a), some off event point sources are visible; these are remnant wake events of sources observed moments before the slew motion through the target field. Some additional mount motion is visible in the accumulated event frame of Figure \ref{fig:medium_accumulated_and_compensated} (a). This slight motion is likely due to wind, but it does not appear to have significantly affected the performance of the motion compensation in Figure \ref{fig:slow_accumulated_and_compensated} (b). 

\subsection{Source Finder and Calibration Performance}

\begin{figure}[]
    \centering
    \subfloat[]{\includegraphics[width=\textwidth]{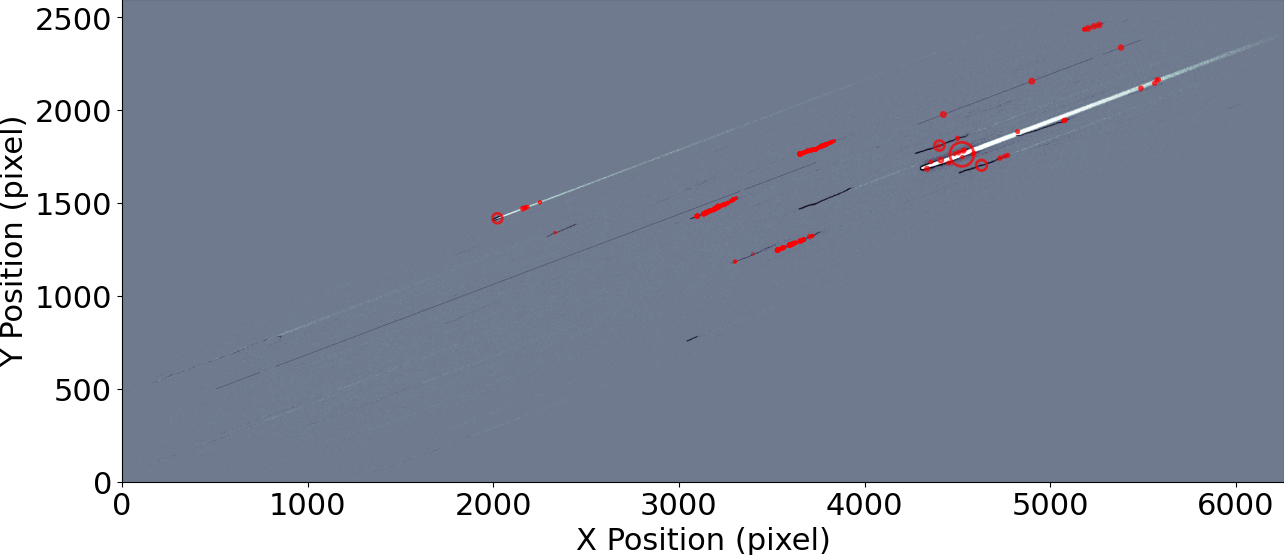}}%
    \qquad
     \subfloat[]{\includegraphics[width=\textwidth]{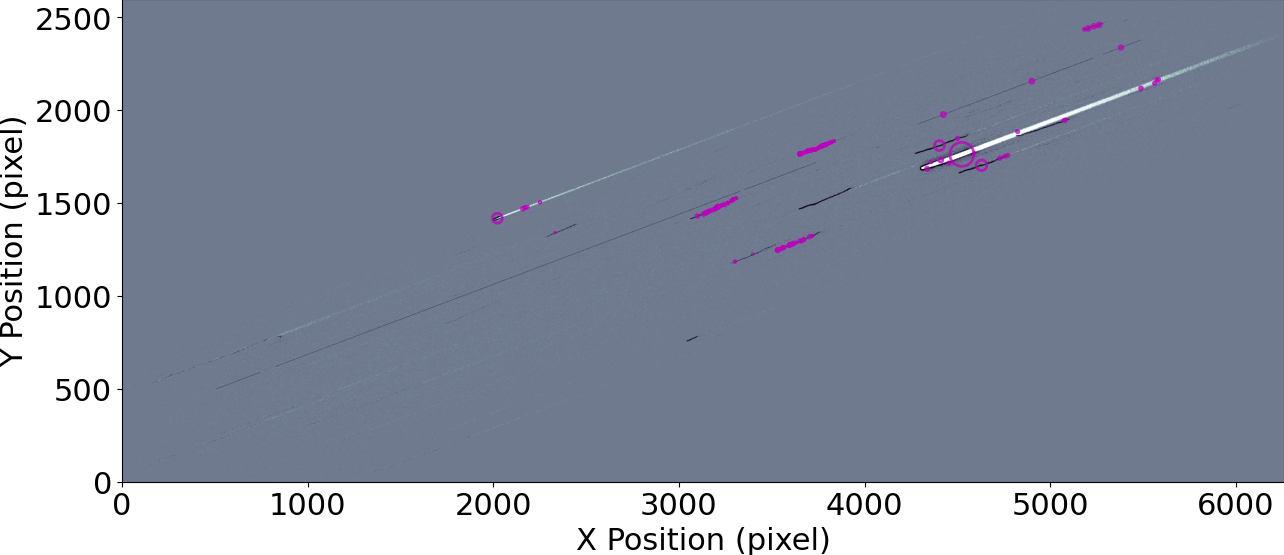}}%
    \caption{Raw source finder output of the field 0 high-speed (0.5 degrees per second) star map with detected sources circled in red (a), and the filtered source detections (magenta) (b). Here, the final output is filtered, but not associated to any astrophysical sources as no astrometric solution could be found for the field. As a result, many erroneous source detections are present in (b), since the calibration solution could not be used for final source filtering.}%
    \label{fig:source_finder_and_calibration_f0_0.5}%
\end{figure}

\begin{figure}[]
    \centering

    \subfloat[]{\includegraphics[width=0.95\textwidth]{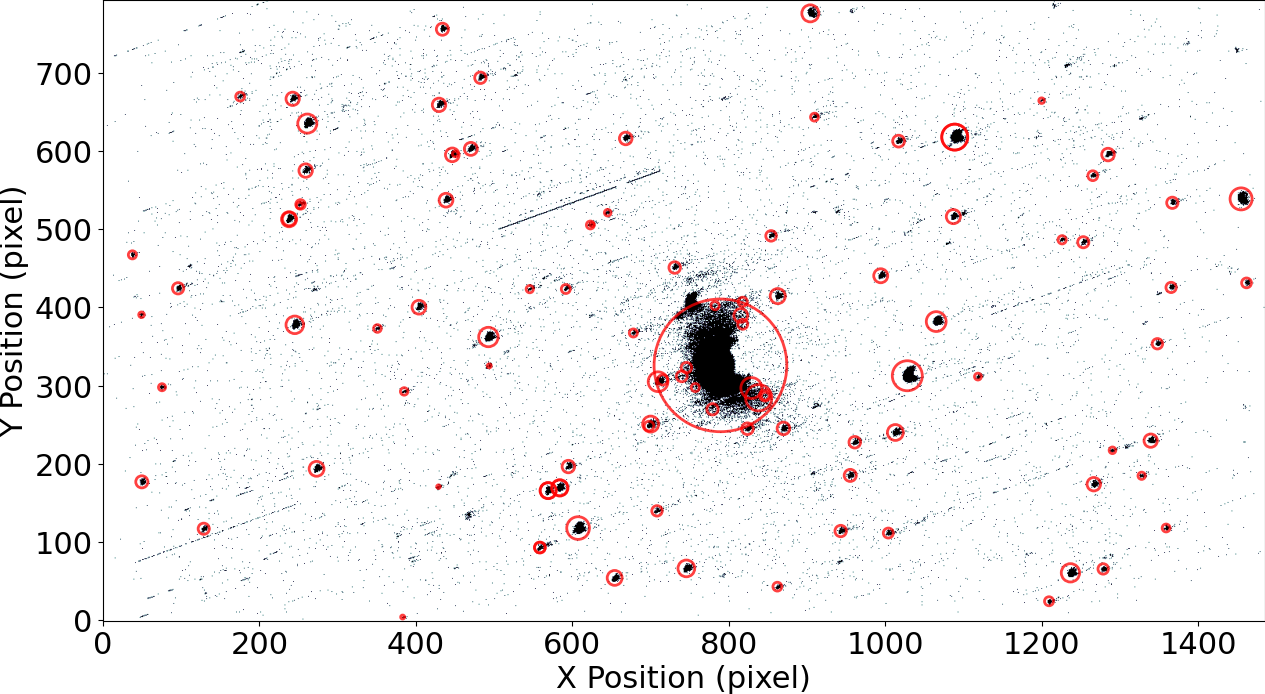}}%
    \qquad
    \subfloat[]{\includegraphics[width=\textwidth]{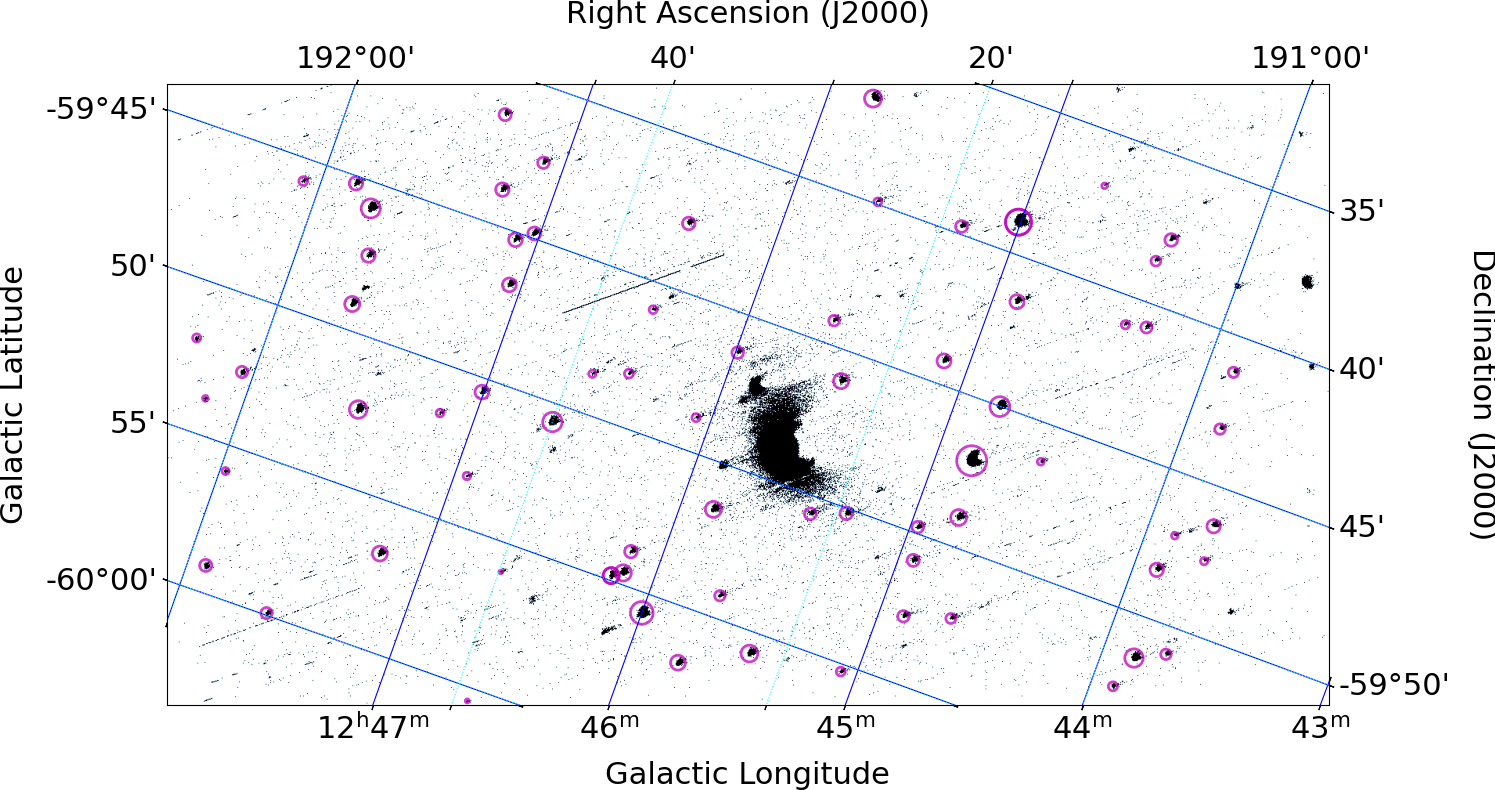}}%
    \caption{Raw source finder output for the field 0 recording without any off-events in the medium speed star map with detected sources circled in red (a). In (b), an astrometric solution is found, and the field is projected onto the WCS coordinate frame. Detected sources successfully associated with a catalogued astrophysical source are circled in magenta. Sources in this field are less numerous than in the star field using both on and off-events. However, fewer spurious false positives are detected in the raw source finder output due to significantly fewer wake events.}%
    \label{fig:source_finder_and_calibration_mono_events}%
\end{figure}

The \ac{EB} source finding algorithm proposed and developed in this paper is demonstrated to successfully detect and localise sources in star maps across multiple fields and slew speeds. In these star maps, the raw output of the source finder is capable of detecting faint point sources. This performance demonstrates the proposed system is largely robust to hot pixel streaks and isolated noise. Particularly in the high-speed star map (Figure \ref{fig:source_finder_and_calibration_f0_0.5}), wake events produce many false positives. When comparing the star maps generated using on events only (Figure \ref{fig:source_finder_and_calibration_mono_events}), fewer noisy regions are detected as sources, but fewer sources are detected overall. The position accuracy of the source finder is high enough to perform a plate solve on the field to calibrate the motion compensated star map to associate the detected sources with real-world astrophysical objects. 

\begin{figure}[]
    \subfloat[]{\includegraphics[width=0.95\textwidth]{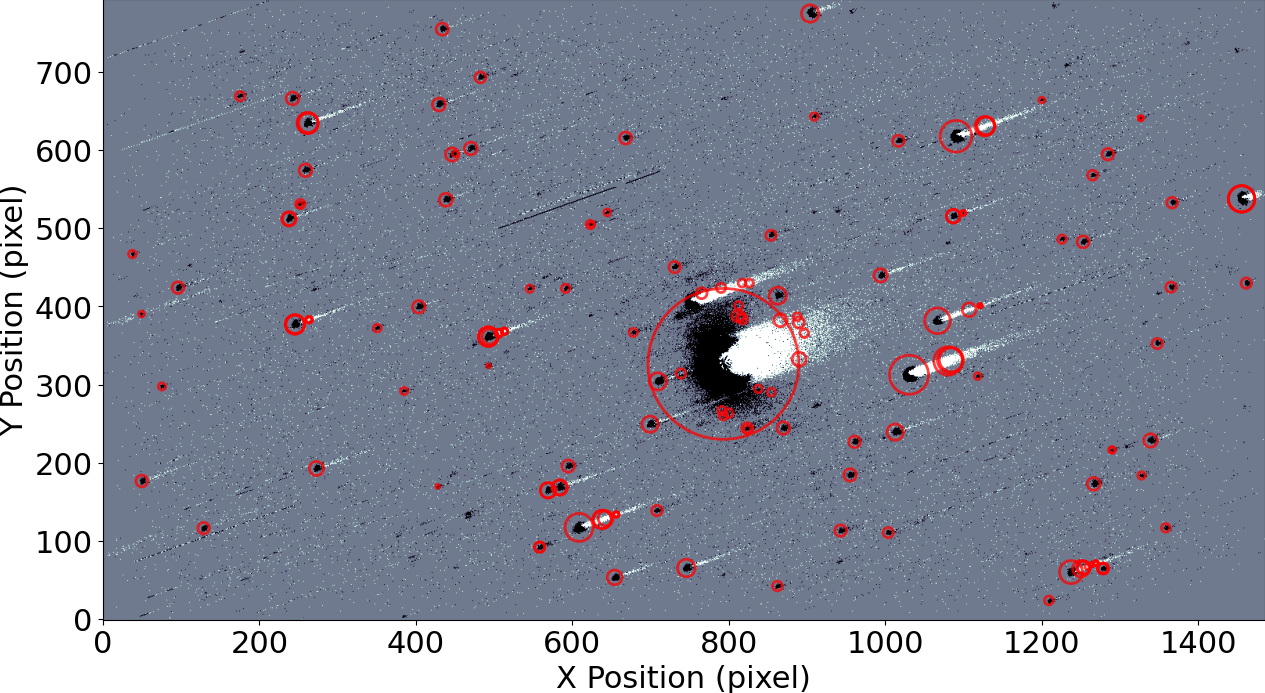}}%
    \qquad
    \subfloat[]{\includegraphics[width=\textwidth]{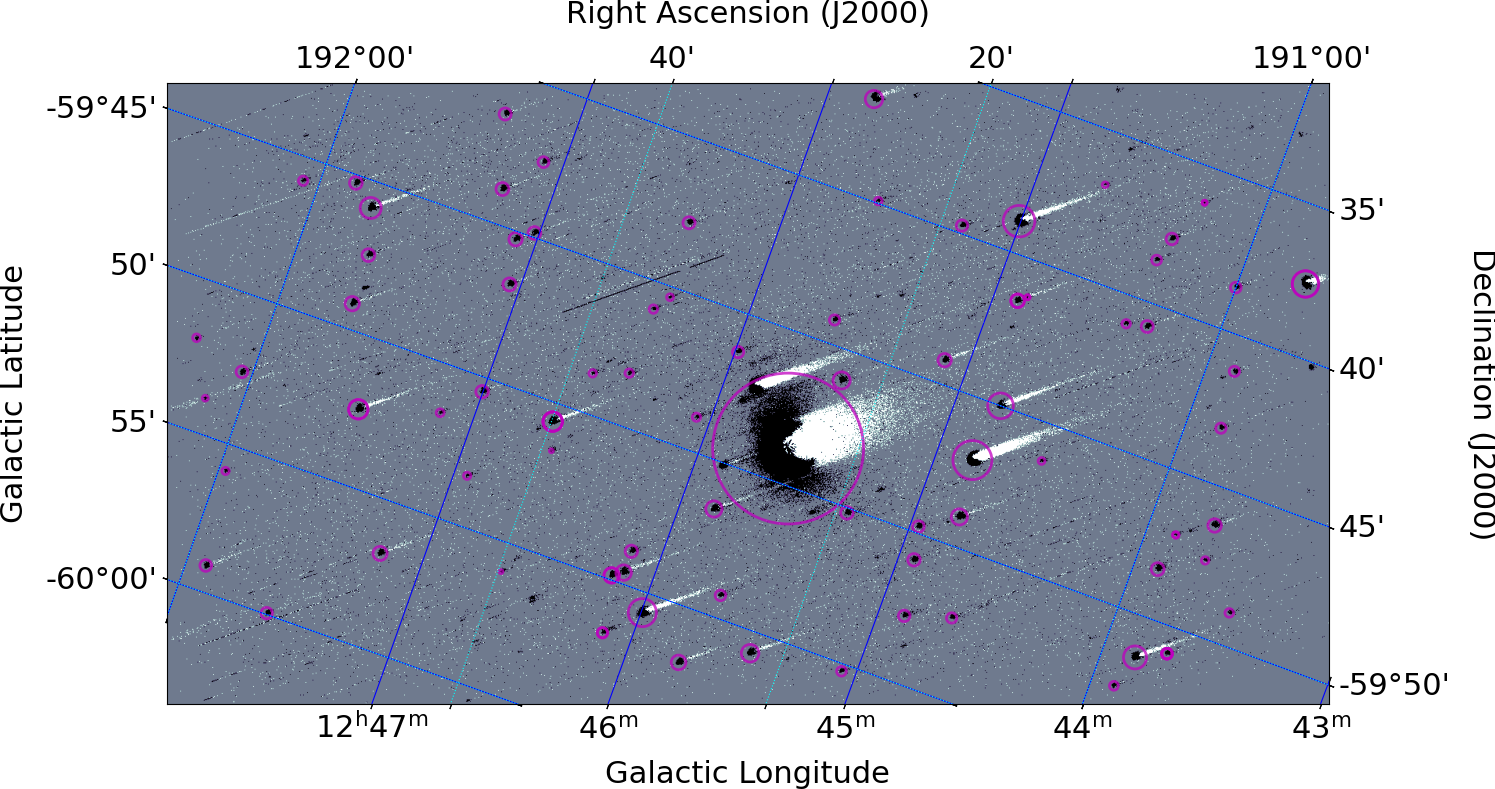}}%
    \caption{Raw source finder output of the field 0 medium speed (0.0015 degrees per second) star map with detected sources circled in red (a). In (b), an astrometric solution is found, and the field is projected onto the WCS coordinate frame. Detected sources successfully associated with a catalogued astrophysical source are circled in magenta. Some sources too near Mimosa were erroneously filtered.}%
    \label{fig:source_finder_and_calibration_f0_0.015}%
\end{figure}

\begin{figure}[]
   \subfloat[]{\includegraphics[width=0.95\textwidth]{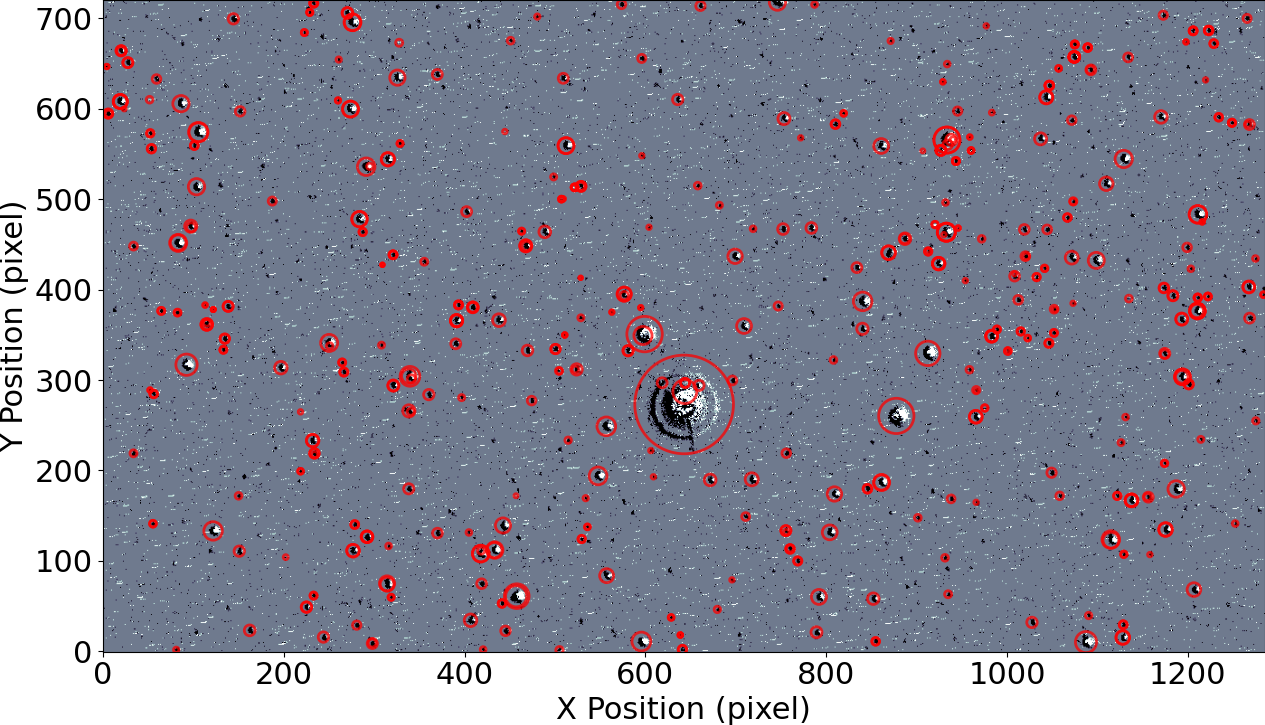}}%
   \qquad
   \subfloat[]{\includegraphics[width=\textwidth]{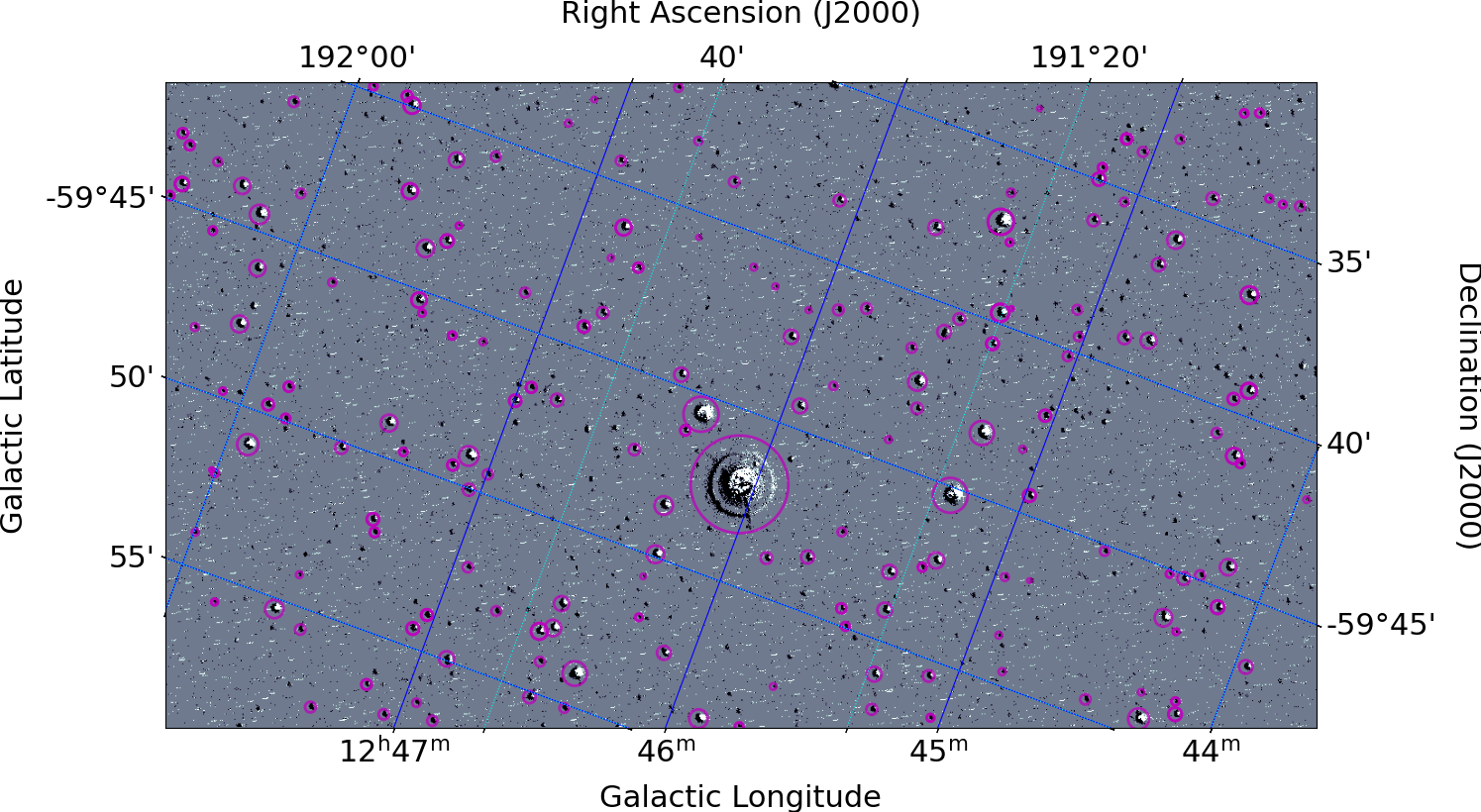}}%
    \caption{Raw source finder output of the field 0 at the lowest speed (0.000488 degrees per second) star map with detected sources circled in red (a). In (b), an astrometric solution is found, and the field is projected onto the WCS coordinate frame. Detected sources successfully associated with a catalogued astrophysical source are circled in magenta. Field 0 at this speed is shown to contain a vast number of detectable sources.}%
    \label{fig:source_finder_and_calibration_f0_0.0004}%
\end{figure}

\begin{figure}[]
    \centering
    \subfloat[]{\includegraphics[width=\textwidth]{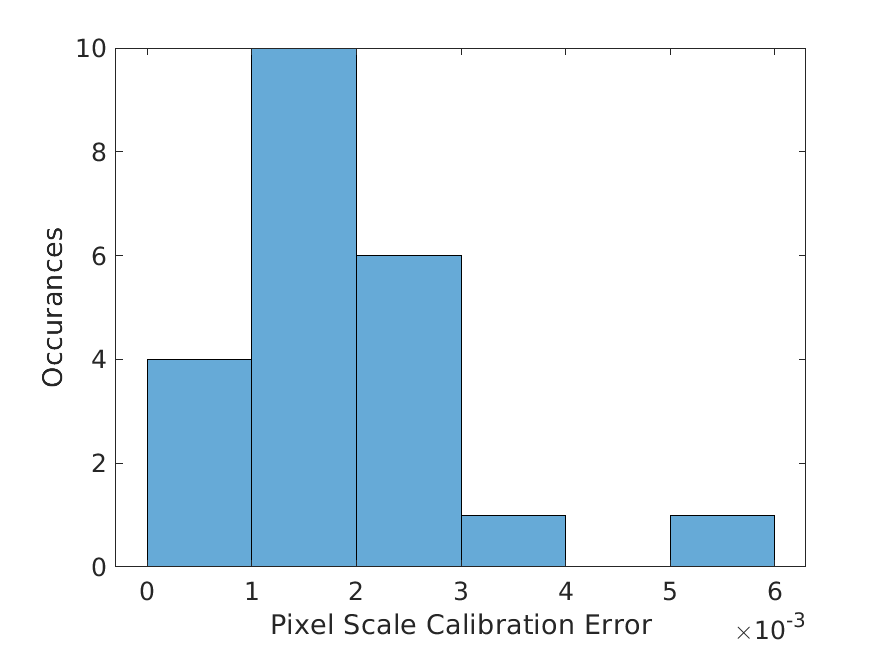}}%
    \qquad
    \subfloat[]{\includegraphics[width=\textwidth]{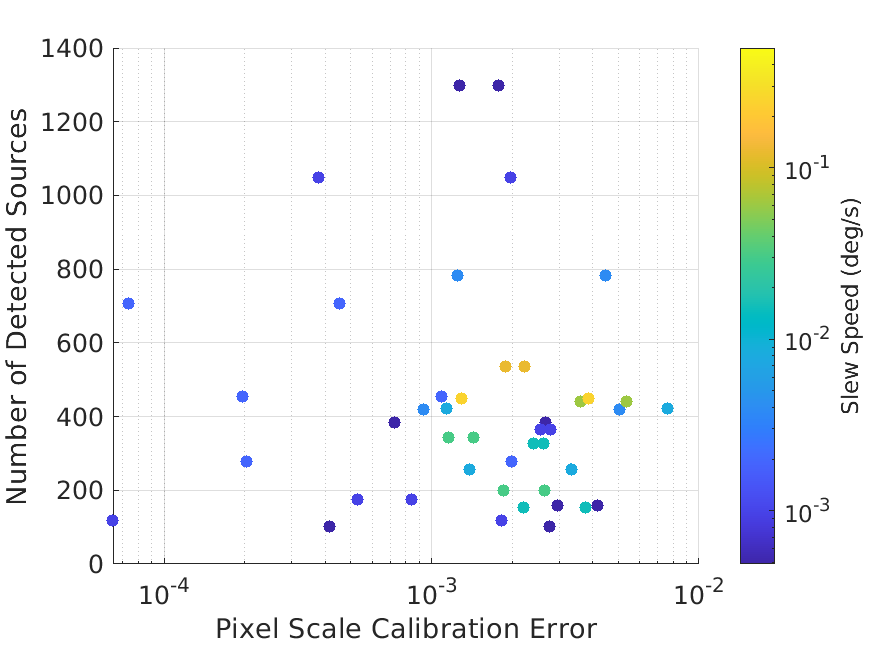}}%
    \caption{In (a), the astrometric calibration error is shown to be consistently low, and well within the imposed $1e^{-2}$ limit. Number of detected sources vs the pixel calibration error, coloured by the slew speed (b), shows higher slew speeds produce higher pixel scale estimation errors, and therefore poorer astrometric solutions, regardless of source count.}%
    \label{fig:astrometric_calibration_performance_pix_scale}%
\end{figure}

An astrometric calibration solution was found for the majority of the fields. Shown in Figures \ref{fig:source_finder_and_calibration_f0_0.5}-\ref{fig:source_finder_and_calibration_f0_0.0004}, the calibration solution can be used to project events as measurements in the pixel frame to positions in the world frame (J2000.0). In Figure \ref{fig:astrometric_calibration_performance_pix_scale} (a), the error between the astrometric estimate of the pixel scale and the ground truth is shown to be low, at an average of 0.0021 arcseconds per pixel, which is an accurate descriptor of the calibration performance. Figure \ref{fig:astrometric_calibration_performance_pix_scale} (b) illustrates a correlation between high-speed slews and poorer pixel scale estimates. In these Figures, the final filtered and associated output source finder is highly robust to noise, wake events and hot pixels, as only sources associated with a catalogued source are flagged as genuine event sources.

\begin{figure}[]
    \centering
    \includegraphics[width=\textwidth]{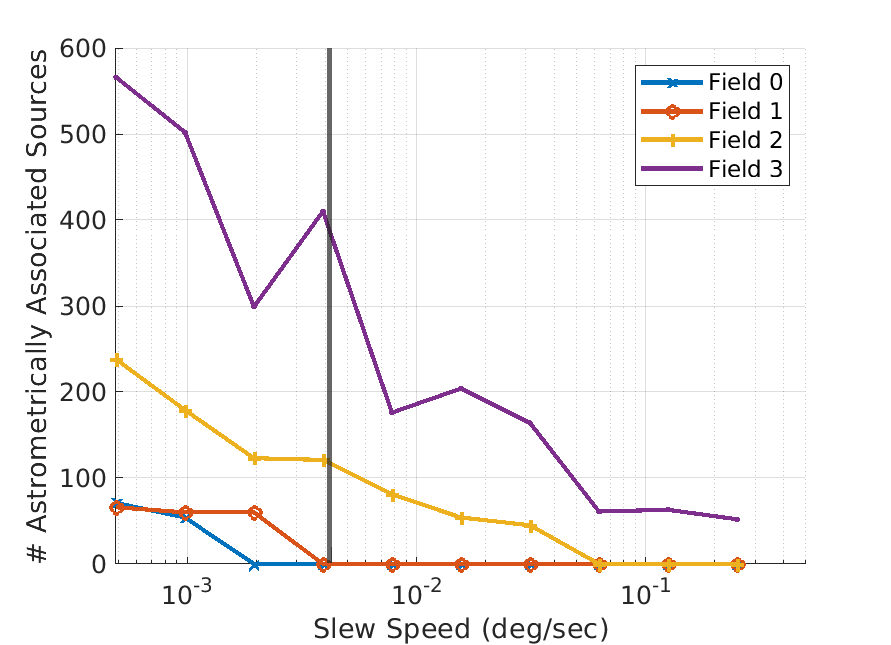}
    \caption{The number of detected sources in each field (0-3), where the number of detectable and associated sources increases across the full range of slew speeds from 0.5 degrees per second at $1200\%$ sidereal rate to 0.000488 degrees per second at $11.5\%$ sidereal rate.}
    \label{fig:detected_sources_each_speed_each_field}
\end{figure}

Figure \ref{fig:detected_sources_each_speed_each_field} demonstrates that the number of detectable sources increases as the speed decreases. This figure also shows the different source counts of each field, where the Mimosa Field 0 has the most numerous. A sharp increase in source count is visible for Field 0 at $0.00075$ degrees per second, where the slew scanned past extra sources which are otherwise outside the \ac{FOV} for the other scans.

\begin{figure}[]
    \centering
    \includegraphics[width=\textwidth]{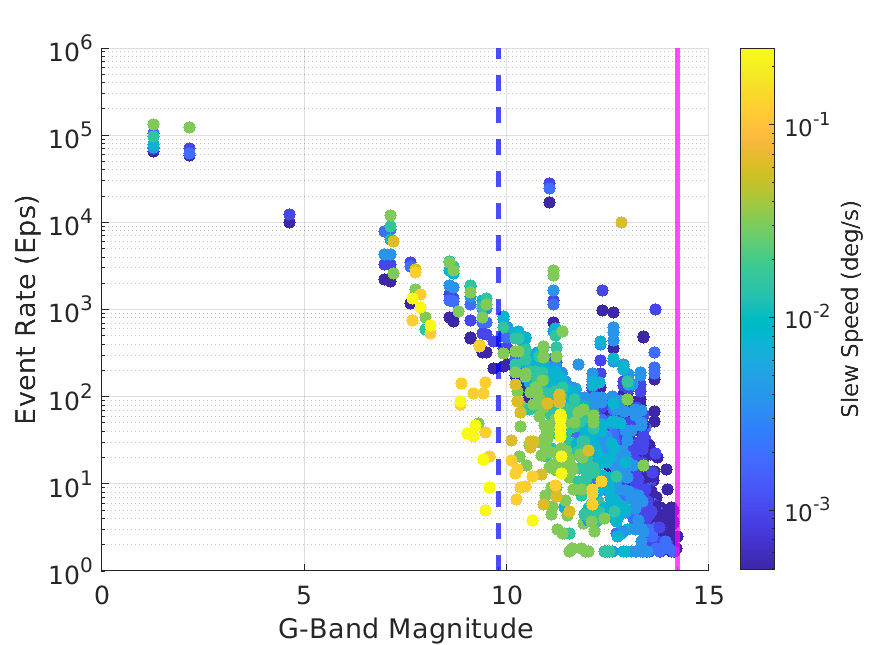}%
    \caption{The event rate vs source brightness produced by the pipeline developed in this paper with the Gen 4 HD with Astrosite-1}%
    \label{fig:comparison_eps_vs_mag_with_cots_paper}%
\end{figure}

\begin{figure}[]
    \centering
    \includegraphics[width=\textwidth]{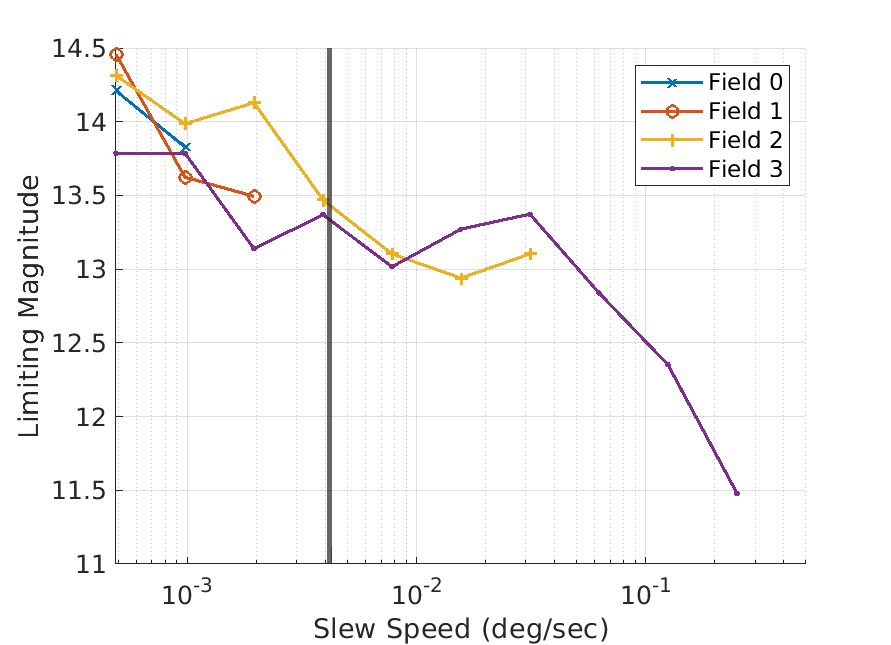}%
    \caption{The limiting magnitude across a range of speeds using the Gen 4 HD and Astrosite-1 with a lower limiting magnitude (sidereal rate shown in a solid black line). Fields without a detection limit occur when no astrometric solution can be found due to insufficient detectable sources at such speeds using the EBC.}%
    \label{fig:mag_limit_per_speed_with_cots_comparison}%
\end{figure}

In Figure \ref{fig:mag_limit_per_speed_with_cots_comparison}, the limiting magnitude of the third generation \ac{ATIS} \ac{EBC} from the previously introduced \cite{mcmahon2021commercial} (Section \ref{sect: measuring_the_sky_intro}) is compared to the limiting magnitude of the Gen 4 HD as determined by the star mapper and source finder system developed in this paper. These results demonstrate that the limiting magnitude of the Gen 4 HD Astrosite-1 setup is significantly higher at magnitude 14.45 during the slowest speed slew, compared to the limit found using the VGA and HVGA ATIS with an 85 mm f/1.4 lens setup of \cite{mcmahon2021commercial}, at magnitude 9.6. The lower observed sensitivity (and therefore higher predicted sensitivity of the Gen 4 HD) is due to both changes in the characteristics of the \ac{EBC}, but also due to \cite{mcmahon2021commercial} using a smaller aperture telescope. This difference in the telescope setup reduces sensitivity due to a smaller aperture. Detected sources using the Gen 4 HD are also shown to have a higher event rate than the previous generation, suggesting the \ac{SNR} of the Gen 4 HD is higher, given the global event rate of the \ac{EBC} is known to be lower \cite{finateu20205}. Furthermore, in Figure \ref{fig:comparison_eps_vs_mag_with_cots_paper}, the Gen 4 HD is capable of detecting fainter sources at higher slew speeds (0.25 degrees per second with a sensitivity up to magnitude 11.4) compared to the previous generation \ac{EBC}. 

Using the output of the source finder and calibrator, the spatio-temporal features and measurement dynamics of event sources are analysed at each slew speed. These dynamics are then related to the properties of the associated astrophysical sources. The most apparent event source features are examined; spatial extent, event rate, on/off event ratio and \ac{COM} offsets. 

\begin{figure}[]
    \centering
    \subfloat[]{\includegraphics[width=\textwidth]{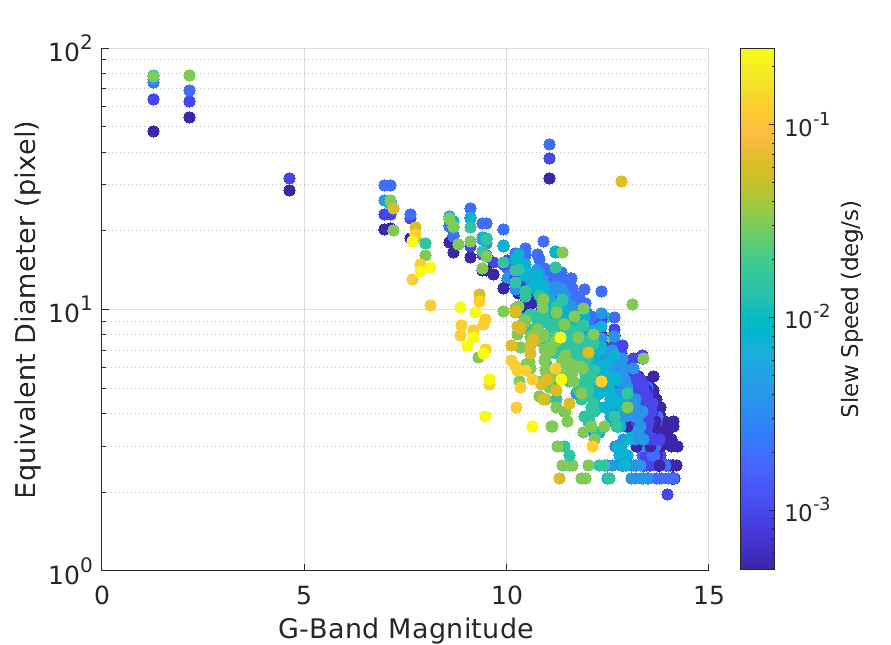}}%
    \qquad
    \subfloat[]{\includegraphics[width=\textwidth]{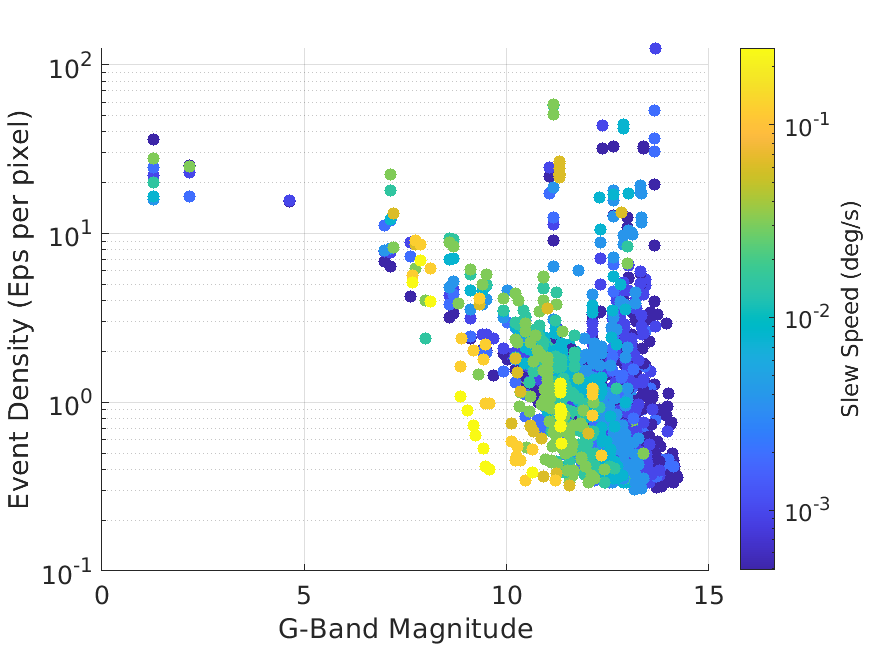}}%
    \caption{Equivalent diameter (smallest bounding circle) (a) and event density vs the magnitude brightness of sources (b), coloured by the slew speed (degrees per second). The magnitude limit decreases with slew speed until a clear lower magnitude limit of 14.45 is visible at the lowest speed. Source spatial extent and density are shown to increase with source brightness. }%
    \label{fig:mag_vs_eps_and_equiv_diam}%
\end{figure}

\begin{figure}[]
    \centering
    \subfloat[]{\includegraphics[width=\textwidth]{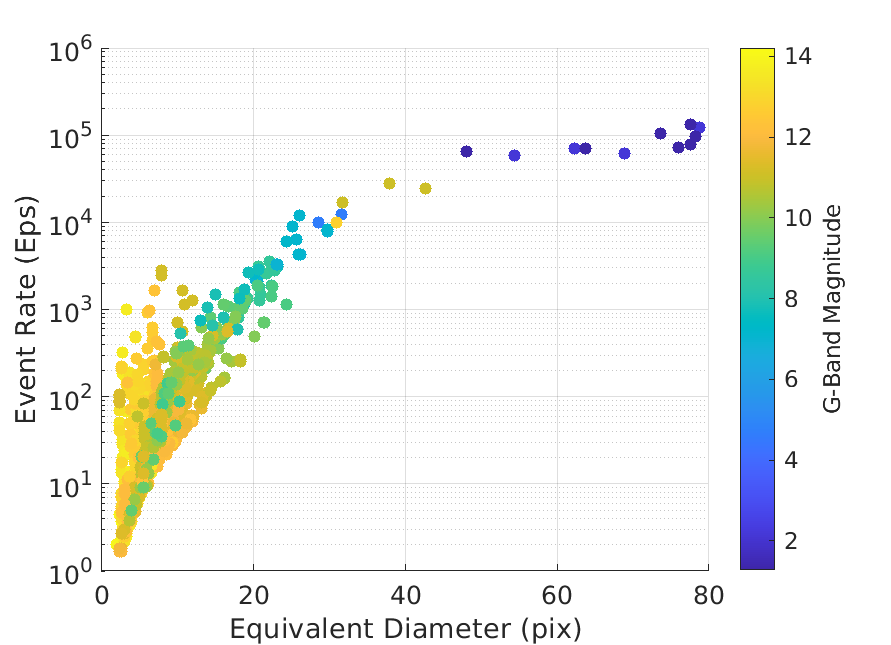}}%
    \qquad
    \subfloat[]{\includegraphics[width=\textwidth]{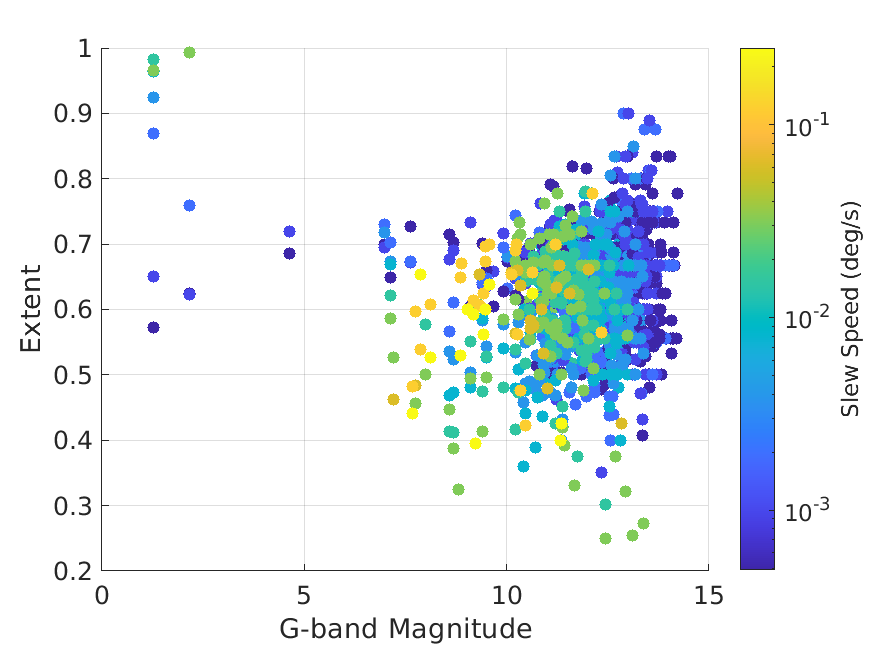}}%
    \caption{Event rate vs equivalent diameter coloured by magnitude brightness (a), showing the event rate of a source increases rapidly with the apparent source size, which is a function of the source brightness. Extent vs magnitude brightness coloured by slew speed (a), where the extent (1 being most circular) shows the brightest sources are the most circular.}%
    \label{fig:calibrated_source_extent_plots}%
\end{figure}

\begin{figure}[]
    \centering
    \subfloat[]{\includegraphics[width=\textwidth]{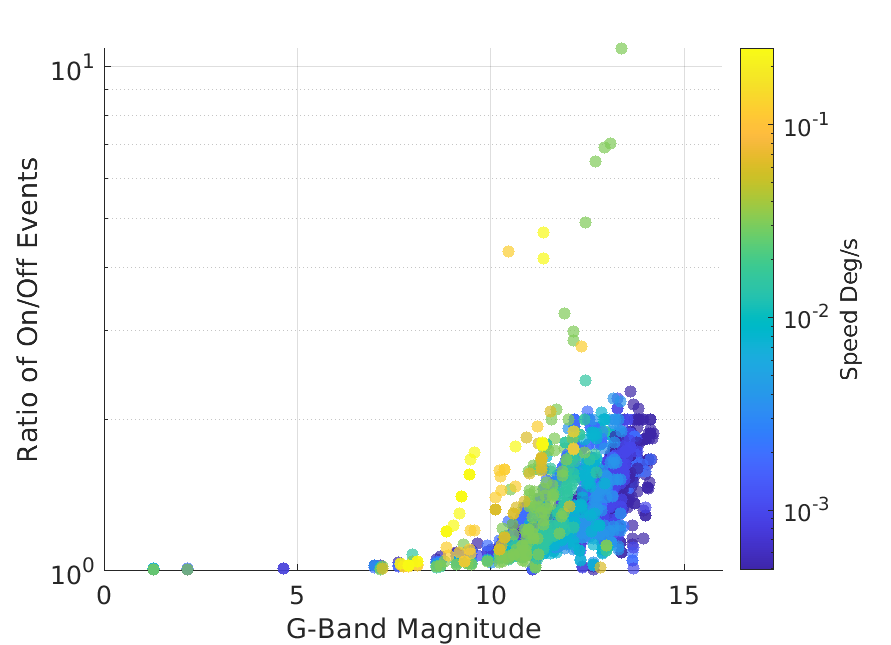}}%
    \qquad
    \subfloat[]{\includegraphics[width=\textwidth]{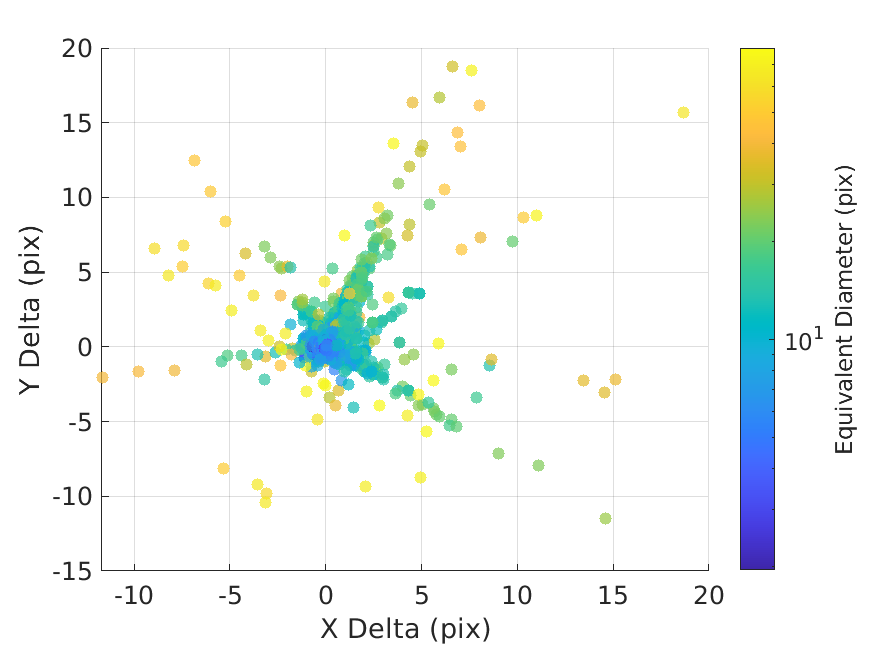}}%
    \caption{On-off event ratio vs brightness magnitude, coloured by slew speed (a), shows that faint sources detected at mid-range speeds produce the most off-events. The off-event count is shown not to exceed the on-event count. In (b), the error between the geometric COM and weighted COM is shown to grow rapidly with the apparent source size.}%
    \label{fig:event_ratio_vs_mg_and_geometric_com_error}%
\end{figure}

As expected, the event rate of sources are shown in Figure \ref{fig:mag_vs_eps_and_equiv_diam} (a) and \ref{fig:calibrated_source_extent_plots} (a) to rapidly increase with the equivalent diameter, where bright sources with a large extent produce orders of magnitude more events than smaller faint sources. However, in Figure \ref{fig:event_ratio_vs_mg_and_geometric_com_error} (b), these bright sources produce fewer off- events than faint sources, regardless of the slew speed. Here, faint sources observed during high-speed slews produce significantly more off-events than sources observed in low-speed slews. Although this trend is certainly influenced by how many off-events are associated with the source by the source finder, as at high-speed, where the off-events form long wake event streaks, the off-events are not associated with the source and are often filtered out. At lower speeds, this is not the case where sources regardless of size have nearly all events successfully associated.  

As shown in Figure \ref{fig:mag_vs_eps_and_equiv_diam} (b), as the source magnitude decreases (brightness increases), the apparent size (estimated using the source equivalent diameter as the minimum diameter of a circle required to enclose the source) of the source increases. Additionally, in Figure \ref{fig:calibrated_source_extent_plots} (b), the circular extent of sources (1 being completely circular and 0 being not at all circular) is higher for sources observed during low-speed slews, where they appear their most circular and symmetrical with comparatively fewer wake events. At higher speeds, sources begin to spread and elongate along the leading edge. 

\begin{figure}[]
    \centering
    \subfloat[]{\includegraphics[width=\textwidth]{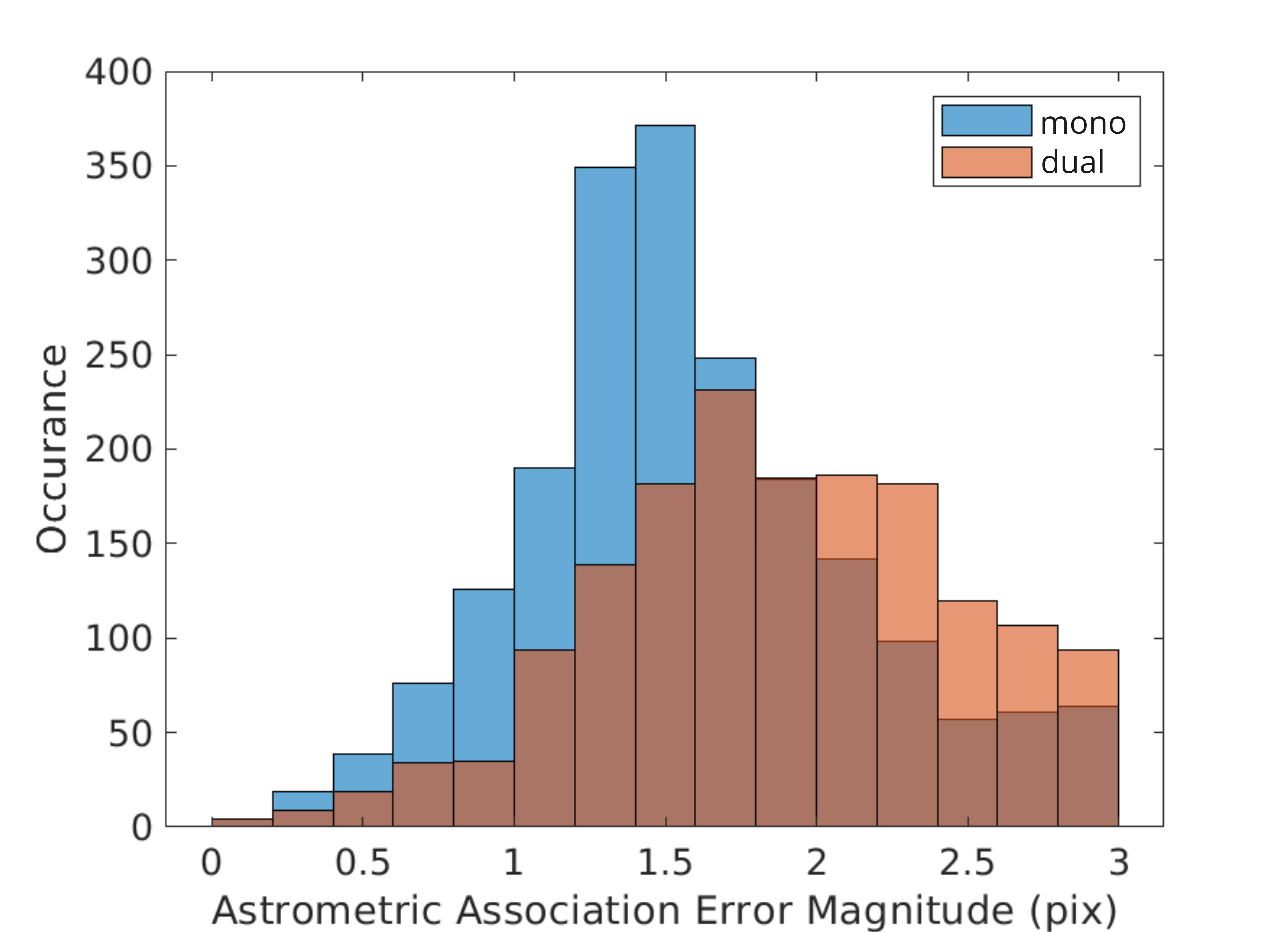}}%
    \qquad
    \subfloat[]{\includegraphics[width=\textwidth]{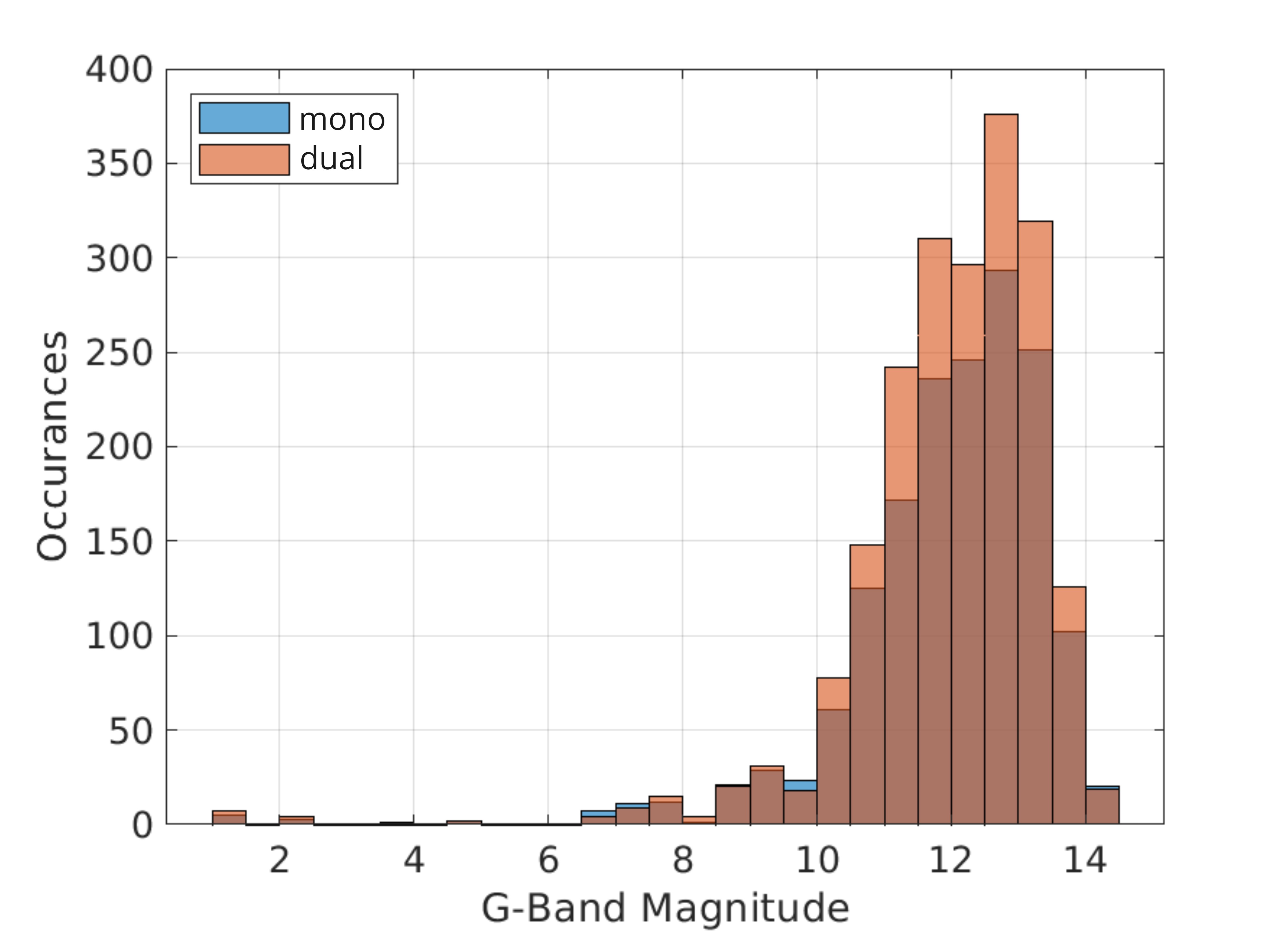}}%
    \caption{Distribution of the astrometric association error between source positions on the calibrated star map and the closest catalogued astrophysical source. By using only on events in the star map (mono), a lower association error is observed compared to using both on and off-events (dual) (a). In (b), the distribution of source brightness detected for mono, and dual polarity star maps suggest the overall number of detectable sources is lower for mono star maps, regardless of magnitude.}%
    \label{fig:source_association_error_whole_field_and_duel_vs_mono_brighness_distrib}%
\end{figure}

\begin{figure}[]
    \centering
    \subfloat[]{\includegraphics[width=\textwidth]{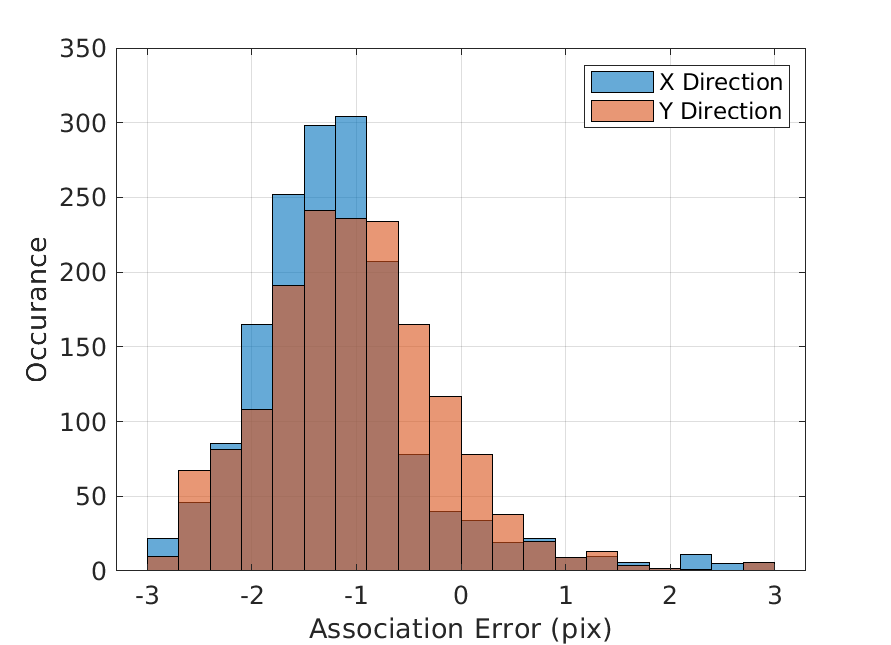}}%
    \qquad
    \subfloat[]{\includegraphics[width=\textwidth]{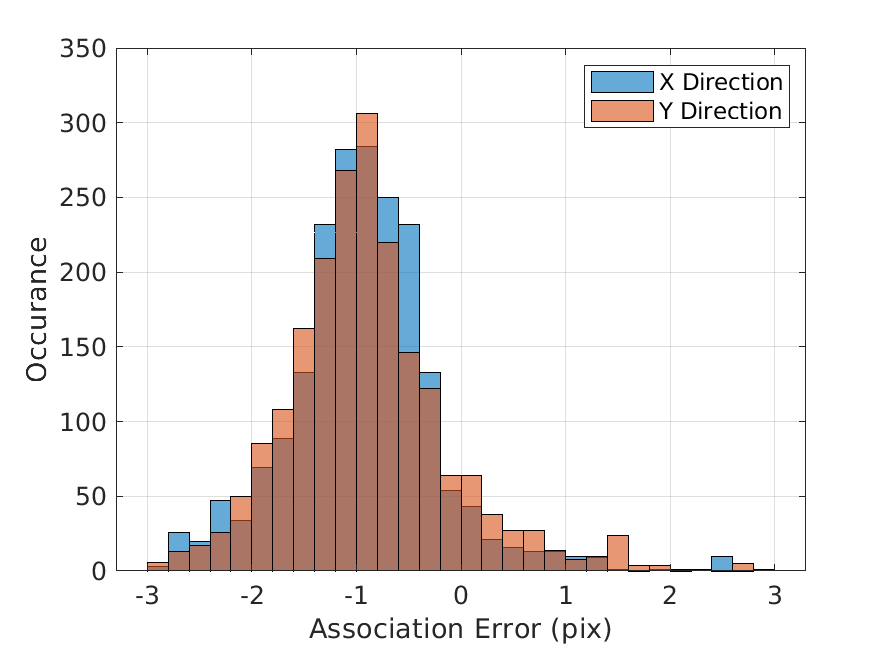}}%
    \caption{Directional association error distribution in the x and y direction for mono polarity (on events only) star maps (a) and dual polarity star maps (b), which shows source COM have a consistent position offset behind the actual source COM. This effect is less pronounced in mono-polarity star maps.}%
    \label{fig:directional_astrometric_position_error}%
\end{figure}

A consistent offset between the associated source and the source finder position is clear in Figures \ref{fig:directional_astrometric_position_error} - \ref{fig:source_association_error_whole_field_and_duel_vs_mono_brighness_distrib}(a). This constant offset suggests the source \ac{COM} is behind the event leading edge. Although the weighted \ac{COM} here is shown to be behind the true \ac{COM}, the error between the geometric \ac{COM} in Figure \ref{fig:event_ratio_vs_mg_and_geometric_com_error} (b) is shown to be high for all but the slowest of slews. The error between the geometric and weighted COM highlights how asymmetrical and unbalanced the source's spatial distribution is, particularly without both on and off-events in low-speed observations.


\begin{figure}[]
   \includegraphics[width=\textwidth]{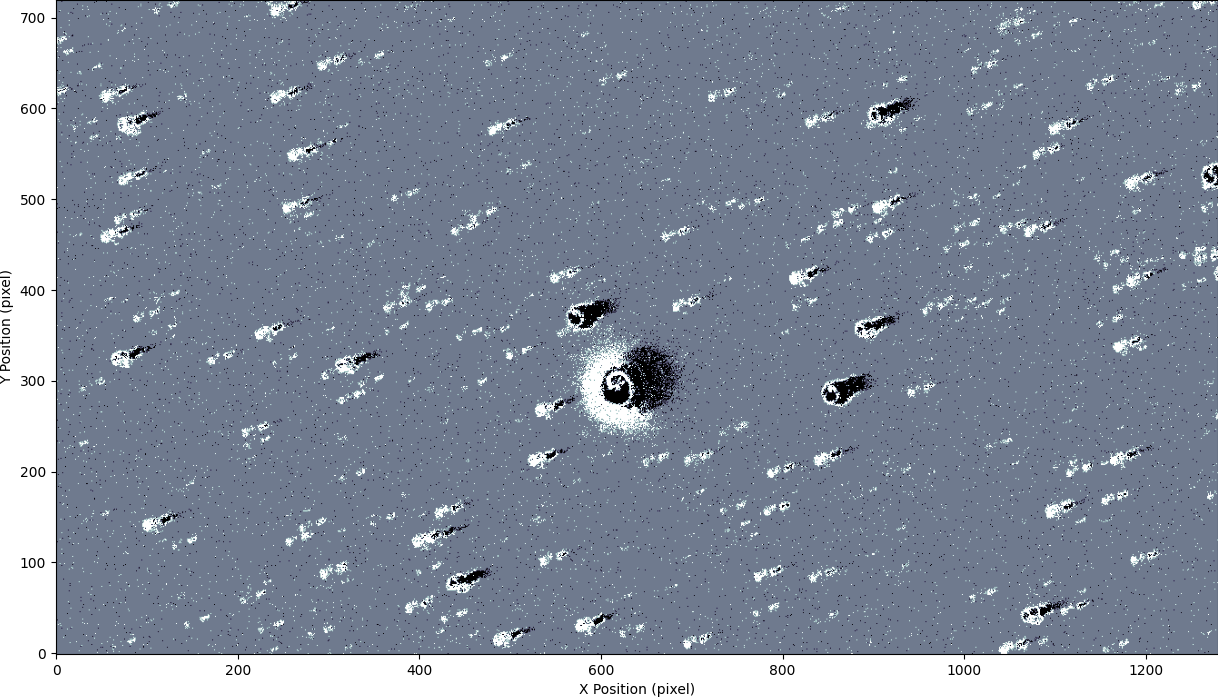}
    \caption{Raw source finder output of field 0 observed in windy conditions where the adverse effects are most apparent at a slew speed of 0.002 degrees per second. Event sources in the star map are highly diffuse and poorly resolved.}%
    \label{fig:windy_accumulated_frame}%
\end{figure}

\begin{figure}[]
   \subfloat[]{\includegraphics[width=0.95\textwidth]{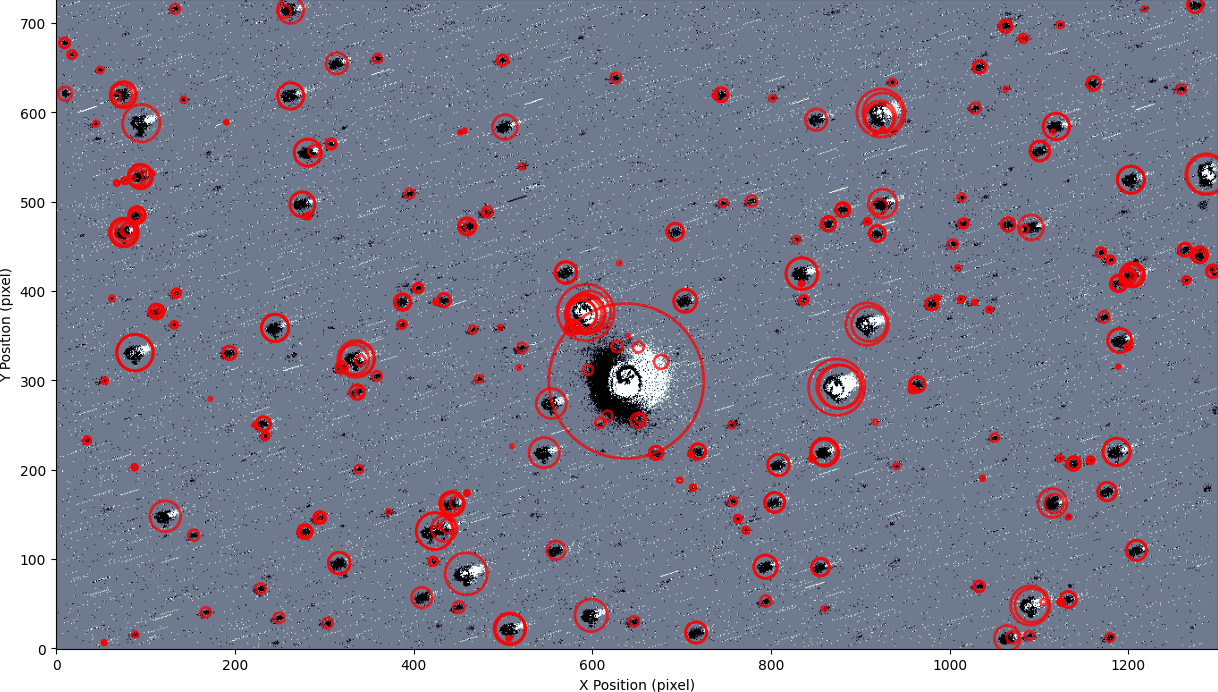}}%
   \qquad
   \subfloat[]{\includegraphics[width=\textwidth]{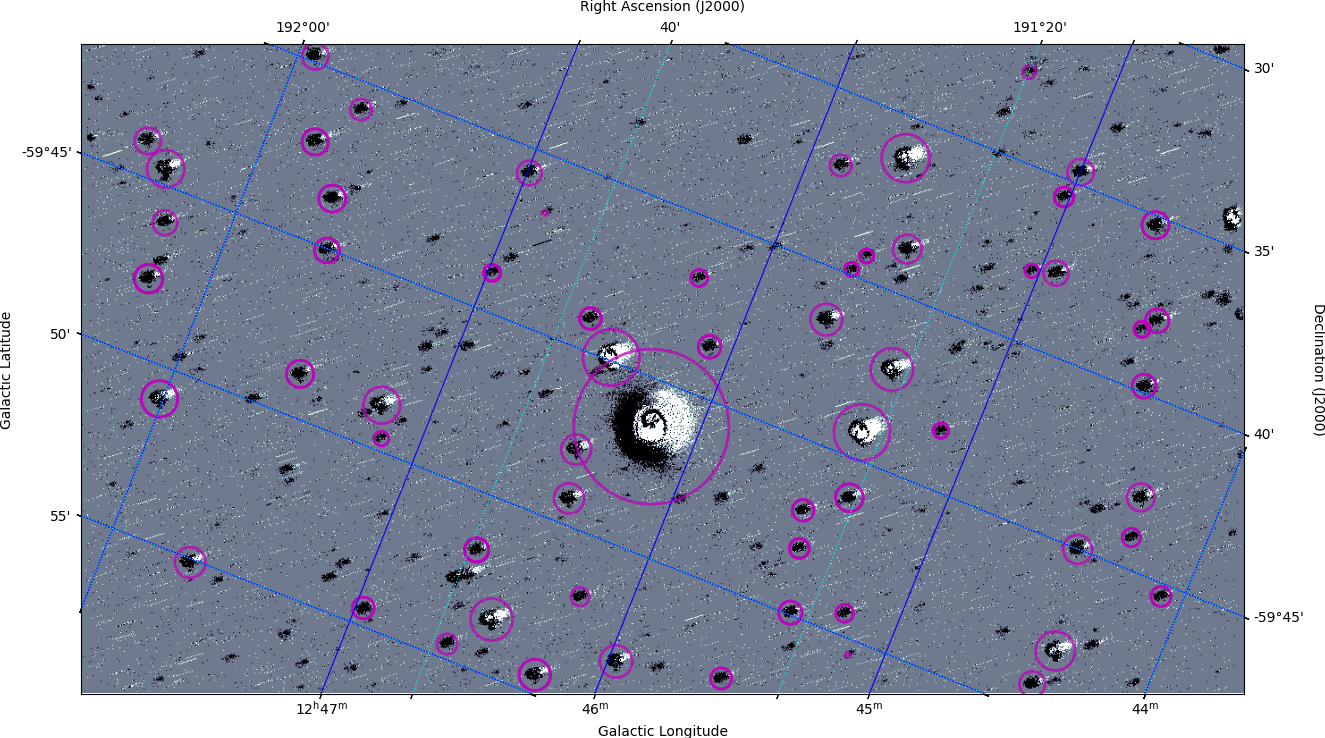}}%
    \caption{Raw source finder output of field 0 observed in windy conditions where the adverse effects are most apparent with a slew speed of 0.002 degrees per second. A star map with detected sources circled in red (a). In (b), an astrometric solution is found, and the field is projected onto the WCS coordinate frame. Detected sources successfully associated with a catalogued astrophysical source are circled in magenta. Field 0 at this speed is shown to contain a vast number of detectable sources.}%
    \label{fig:windy_source_finder_and_calibration}%
\end{figure}

\begin{figure}[]
    \centering
    \subfloat[]{\includegraphics[width=\textwidth]{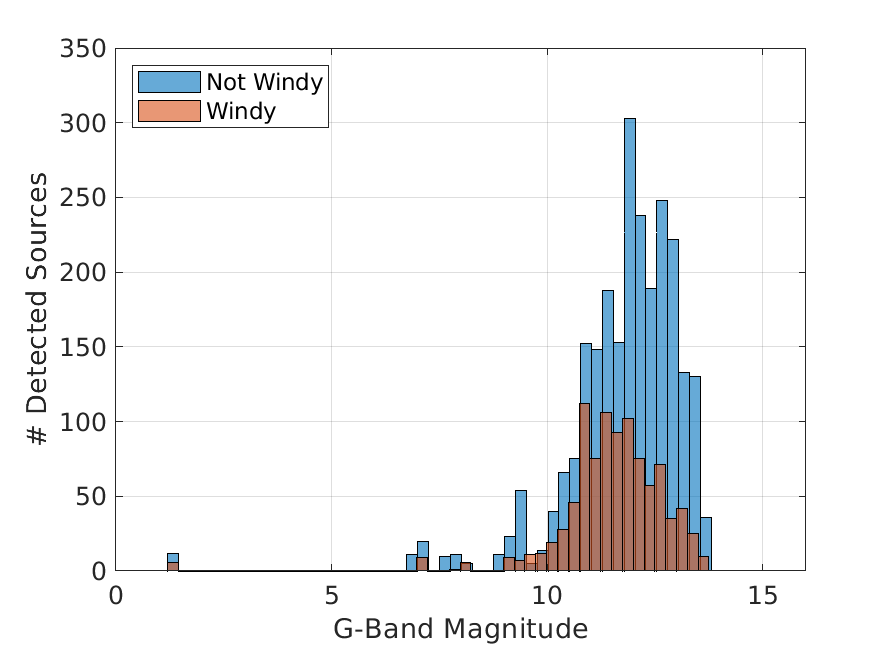}}%
    \qquad
    \subfloat[]{\includegraphics[width=\textwidth]{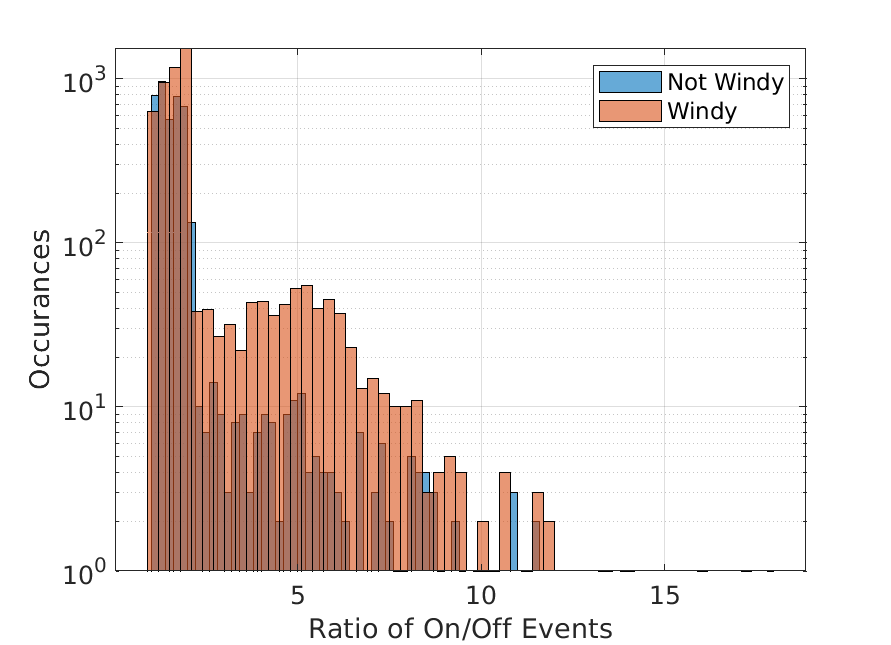}}%
    \caption{A comparison of the distribution between the brightness of detected sources for windy and non-windy observations (a) shows that windy conditions reduce the number of detectable sources. In (b), the on-off event ratio for windy recordings shows sources detected in windy conditions produce significantly more off-events.}%
    \label{fig:windy_histogram_mag_sources_and_event_ratio}%
\end{figure}

\begin{figure}[]
    \centering
    \subfloat[]{\includegraphics[width=\textwidth]{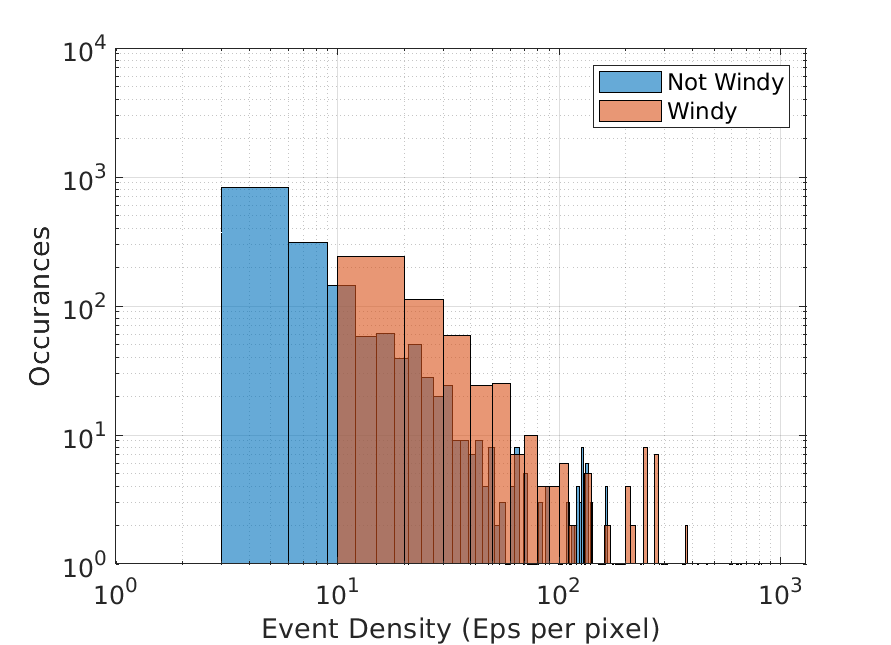}}%
    \qquad
    \subfloat[]{\includegraphics[width=\textwidth]{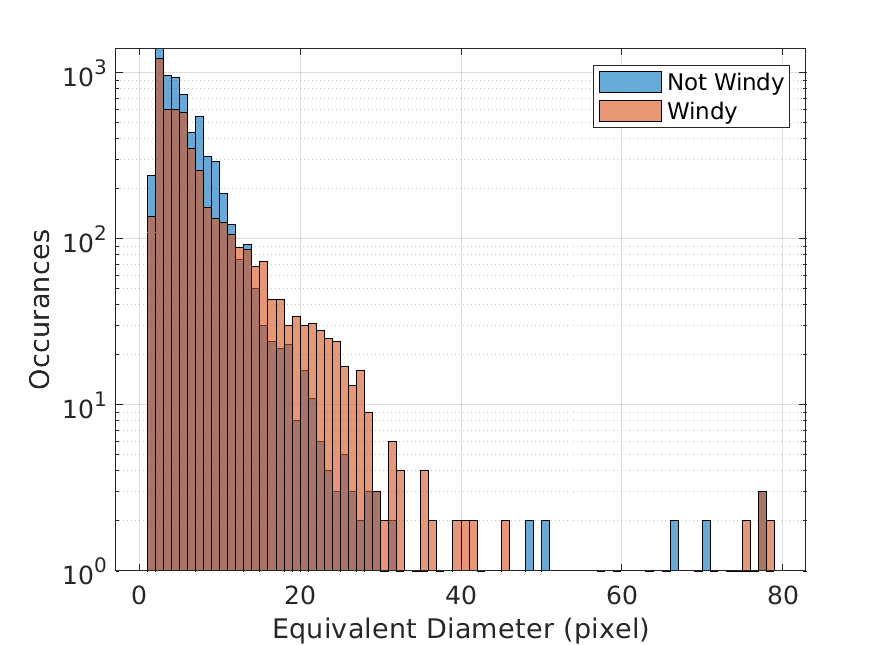}}%
    \caption{Comparisons between the distribution of event density (a) and equivalent diameter (b) for windy and not-windy observations, where the event density and spatial extent of sources are shown to be higher for sources detected in windy conditions.}%
    \label{fig:windy_event_density_and_equiv_diam}%
\end{figure}

On a separate observing night, the speed survey experiments were conducted on Field 0 during windy conditions at an approximate speed of 20~km.h$^{-1}$. Despite the adverse conditions, the star mapping and source finding pipeline was demonstrated to perform robustly to the random motion of the field. As a result, the wind-affected observations were successfully motion compensated to produce star maps with diffuse event sources, as demonstrated in Figure \ref{fig:windy_accumulated_frame}. The source finder also successfully detected these resulting event sources as in Figure \ref{fig:windy_source_finder_and_calibration} (a), which were then used to calibrate the field (b). While the windy observations were successful processed, these observations contained significantly fewer detectable sources across all magnitudes, Figure \ref{fig:windy_histogram_mag_sources_and_event_ratio} (a). Moreover, fewer associated sources were detected, demonstrated in the reduced source detections between the raw source finder output and the calibrated output, where poorer source \ac{COM} estimates failed to produce sufficient source associations. In addition to the reduced sensitivity and poorer \ac{COM} estimates, the spatio-temporal characteristics of the event sources observed in windy conditions differ from those observed without wind. Event sources observed in these conditions produced a higher event rate, event density and equivalent diameter, as illustrated in Figure \ref{fig:windy_histogram_mag_sources_and_event_ratio} (b) and Figure \ref{fig:windy_event_density_and_equiv_diam} (a). These characteristics are consistent with targets undergoing additional mutual motion, causing the oncoming source light to move randomly in the field, increasing the detectable contrast and spatial extent of the \ac{PSF}. Additionally, in Figure \ref{fig:windy_histogram_mag_sources_and_event_ratio} (b), sources observed during windy conditions are shown to produce significantly more off-events.

\section{Discussion}\label{sect:measuring_sky_discusssion}

In these experiments, the performance of the proposed pipeline is successfully demonstrated, and the changing spatio-temporal characteristics of event sources are analysed at varying slew speeds. This system is an accurate and effective solution to source finding and astrometric calibration. It is the first solution in the literature to solely use real-world collected \ac{EB} data in the source finding and astrometric calibration. Beyond use as an analysis tool, the potential role of the pipeline developed in this paper within a fully-fledged \ac{SSA} system is in performing accurate offline conversion of \ac{RSO} state estimates in the image plane to a world coordinate frame.   

Overall, the proposed star mapper and source finder operated successfully and produced a low astrometric calibration offset. Using the specified optical setup, the limiting magnitude of the Gen 4 HD was found to be 14.45, which is largely based on the capabilities of the algorithms used in this paper, and will improve in future work with several enhancements and as the overall space imaging system develops. Sources increase in apparent size and event rate as brightness increases (magnitude decreases). Faint point sources produce more off-events than bright sources. Wake events produced by off-events are most prominent at high speeds and are difficult to process. Despite the challenges involved in processing off-events, their use improves source finder performance in terms of source count and \ac{COM} estimation accuracy. While this pipeline operates robustly to windy conditions, these conditions reduce the number of detectable sources and the sensitivity limit while increasing the event density, the apparent spatial extent, and with more off-events. The difference between the geometric centre of sources and the weighted centre is non-negligible, and a consistent offset is detectable due to wake events. There is a significant offset between the geometric \ac{COM} and weighted \ac{COM} of sources, which increases with apparent spatial extent. A consistent offset is observed between source weighted \ac{COM} and the \ac{COM} of the associated underlying astrophysical source, where the event source \ac{COM} appears to proceed the catalogued source by approximately 1.5 pixels.

In these results, the number of detectable sources increased as the slew speed lowered. However, an upper bound on the trade-off between the sensitivity limit and slew speed was not observed. Additional lower-speed scanning slews are required in future work to determine the point in which the slew speed becomes too low. At this point, event sources will produce too little contrast which results in a poorer sensitivity. The low-level hardware control of the mount and the telescope mount support must be improved to ensure the telescope is sufficiently stable to produce such low-speed mount slews with precise motion only in the intended direction of travel. Although the mount is instructed to move at a constant speed during data collection, additional uncompensated motion can introduce mutual motion and vibration. In future work, it will not be assumed that a constant velocity field velocity occurs throughout the observation. Instead, a \ac{MTT} algorithm (as opposed to the \ac{STT} approach used in this paper) will be used with a shorter time interval to compensate for any other mutual motion adequately.

In these experiments, off-events produce numerous spurious events and artefacts. This phenomenon caused the astrometric association error between sources and catalogued astrophysical objects to be higher in star maps with only on-events. However, including both event polarities improves the symmetry of event sources in low-speed observations, reduces the source measurement's complexity, and makes for easier tracking. Furthermore, with both polarities, the total source count across the brightness range increases as event sources have an overall higher event rate. This increase could be viewed as an increase in the \ac{SNR}; however, further investigation must be conducted to precisely classify events as signal or noise events.


The high temporal resolution of the \ac{EBC} is exemplified by its robust performance in the presence of wind. While the pipeline performed well in adverse windy conditions, a reduction in the number of detectable sources was observed due to several factors. Primarily, the windy conditions cause the optics to move, which spreads the oncoming light over a larger region on the image plane. This movement negatively affects the \ac{COM} estimate accuracy and the detection limit of the \ac{EBC}. This negative effect is similar to the problems found while observing using narrow \ac{FOV} telescope imaging systems where the sensitivity can be reduced as contrast is spread across too many pixels \cite{ralph2019observations}. However, the reduced source count is also a consequence of the star mapper failing to compensate correctly for the varying field motion to create compact sources. Comparison between fields observed during windy conditions suggests that mutual motion between the \ac{EBC} and mount is required to overcome the perfect tracking problem to generate enough contrast to detect sources. Although, these experiments show that if this motion is random and too great, the sensitivity limit increases, and sources become more difficult to detect as they become more diffuse.  

The high dynamic range of the \ac{EBC} is apparent while observing fields containing both bright and faint sources, which would typically not be visible in a conventional \ac{CCD} which may saturate under similar conditions. Using the specified optical setup, the limiting magnitude of the Gen 4 HD was found to be 14.45. This limit is largely based on the capabilities of the algorithms used in this paper. Furthermore, this limit can be improved with alternate optics or improved star mapping, source finding and calibration algorithms, which is the focus of future work. Compared to the results from \cite{mcmahon2021commercial}, there is a drastic improvement of the Gen 4 HD over the previous generation \acp{EBC}, which were observed to have a limiting magnitude of 9.6. The improved performance of the newer generation sensor is largely due to the increased pixel resolution and back-side illuminated pixel design for a larger fill factor and lower photon noise. As discussed, the limiting magnitudes of the previous generation \ac{EBC} are impacted by the author's use of a smaller aperture. In the slow-speed slew experiments, the Airy disk of bright sources is visible using the Gen 4 HD due to the increased sensitivity compared to the previous generation sensors. As space imaging studies using the Gen 4 HD have yet to be published, this is the first known observation of an Airy disk pattern surrounding bright astrophysical sources using an \ac{EBC}. 


Although this system is accurate, performs as intended, and contains well-optimised components, it does not perform in real-time. In this case, calibration is purely a post-processing step in converting pixel space measurements to \ac{WCS} and therefore, it does not need to run in real-time, online or parallel with an \ac{EB} \ac{RSO} tracker such as \ac{FIESTA} \cite{ralphreal}. Instead, it can run offline after observations to produce accurate measurement sets for ingestion into a relevant mission system. Pixel positions and coarse angular measurements are sufficient for all other tasks, such as closed-loop tracking. Within this pipeline's filtering and clustering stages, there are several parameters, such as the $\epsilon$ distance in \ac{DBSCAN}, which can consume excess memory if dense regions of events are split into smaller regions that could be detected as individual sources. For example, \ac{DBSCAN} with too many input clusters and a large $\epsilon$ ($\ge 2$) requires a large adjacency matrix to be calculated and stored at run time which can be memory intensive. 

The significant difference between the source weighted \ac{COM} and geometric \ac{COM} reinforces the difficulty of \ac{EB} \ac{SSA} tracking tasks. Consequently, simple shape/kernel-based or feature template tracking approaches that are common in the \ac{EB} vision literature will likely be unable to accurately estimate the \ac{COM} of a source. The apparent offset between the astrophysical source and the weighted \ac{COM} implies that the \ac{COM} state may not be estimated by the position of events alone, as events appear along the leading edge of the target and are not guaranteed to represent the source \ac{COM}. These results show that the majority of events occur on the leading edge, meaning the weighed \ac{COM} proceeds the true source \ac{COM} by a significant offset of approximately 1.5 pixels (or 2.4 arc seconds). This consistent offset further motivates the need for an \ac{ETT}, which assumes that a target has a non-negligible spatial extent, as opposed to more traditional point target tracking. Additionally, since the \ac{COM} error increases for larger extended sources compared to point sources, an \ac{ETT} tracking system would ideally be developed. Such a tracker would be capable of dynamically estimating the \ac{COM} offset and true target \ac{COM} based on the target extent. However, \ac{ETT} algorithms are complex and challenging to implement.


These experiments demonstrate the differences in the measurement properties between compact faint point sources and bright extended sources. Bright sources will appear with a significantly larger spatial extent and higher event rate, producing fewer off-events than fainter point sources. The difference in the event rate and size of bright sources suggests that a method to filter or mask bright objects may be required to prevent saturation or drastic changes to processing times while they are within the \ac{FOV}. Techniques to better estimate the \ac{COM} of these sources must be conducted, especially if the \ac{EBC} is to be used in a large aperture narrow \ac{FOV} imaging systems, where their apparent size could be significantly higher, and even faint sources will appear with a large extent. In this project, the effect of the \ac{EBC} bias settings is not analysed. These biases control the overall behaviour of the \ac{EBC}, such as the on/off event threshold. Future work will be focused on dynamically tuning these biases in response to the scene dynamics or \ac{SSA} task.


\begin{table}[]
\centering
    \resizebox{\columnwidth}{!}{%
    \begin{tabular}{|c|c|c|c|}
    \hline
    \textbf{Slew Speed} &
      \textbf{\begin{tabular}[c]{@{}c@{}}Limiting \\ Magnitude\end{tabular}} &
      \textbf{\begin{tabular}[c]{@{}c@{}}Time to Scan \\ 1 deg$^2$\end{tabular}} &
      \textbf{\begin{tabular}[c]{@{}c@{}}Time to Scan \\ GEO belt\end{tabular}} \\ \hline
    \begin{tabular}[c]{@{}c@{}}0.0005 deg/s\\ (1.8 arcsec/s)\end{tabular} &
      14.45 &
      59 minutes &
      100 hours \\ \hline
    \begin{tabular}[c]{@{}c@{}}0.5 deg/s\\ (1800 arcsec/s)\end{tabular} &
      11.4 &
      3.55 seconds &
      6 minutes \\ \hline
    \end{tabular}
}
\caption{Contextualising the high capacity observing capabilities of the 4th Gen HD, where the pixel scale is 1.584 arcsec per pixel and where the GEO belt is 180 degrees $\times$ 0.5632.}
\label{tab:chapter_2_scanning_speeds_vs_sensitivity}
\end{table}

 The relationships found between the event rate and the equivalent diameter of event sources could be used to estimate target brightness variations in short time intervals, which is highly challenging with current conventional sensors. Estimating these photometric properties at high speed is advantageous as such a technique could capture valuable information for the difficult task of \ac{RSO} characterisation. In Table \ref{tab:chapter_2_scanning_speeds_vs_sensitivity}, the track capacity (the number of targets observable/trackable within a given period of time, usually 1 `night') is contextualised at the limits of the slew speed range examined in this paper. \ac{GEO} belt scanning is a common \ac{SSA} task where all \ac{GEO} are sequentially observed along the orbit, and un-cued or lost \ac{GEO} can be discovered. These slew speed ranges suggest that a scanning observation could be performed at using a 100-hour scan for an emphasis on sensitivity at 0.000488 degrees per second, or a rapid scan with an emphasis on speed can be performed in 6 minutes at 0.25 degrees per second. A similar set of observations with the same compromise can be performed for a region scan with a 1 deg$^2$ area at an arbitrary sky region in 59 minutes and 3.55 seconds, respectively. Within these scanning speeds, an \ac{EB} space imaging system could be used for rapid surveillance of bright targets for the whole \ac{FOR} (the entire observable sky of the system, as opposed to the \ac{FOV} which is the instantaneous coverage of the sensor). Overall, the wide operating range and potential for rapid characterisation is a unique capability in space imaging, which further motivates the use of \ac{EBC} in \ac{SSA}. 
 
Although there is a focus in the literature and this paper on developing software solutions to calibration, the ideal solution may lie in specific \ac{EB} space imaging hardware for \ac{EB} imaging. Aside from the challenges of calibrating an \ac{EB} or generating a pointing model using \ac{EB} data (inverse kinematic model between sky coordinates and joint motion in the mount), the central dilemma is the un-synchronised and uncoordinated clocks and difference in the temporal resolution between the telescope mount position (given by the encoder read-out) and the \ac{EBC}. 


\section{Conclusion}

This paper develops a star mapping and source finding algorithm to create resolved images of event sources at varying speeds to calibrate the Gen 4 HD \ac{EBC} for accurate measurement acquisition for real-world space imaging tasks. This star mapping and calibration pipeline is the first automatic and fully \ac{EB} system in the literature, which has been built and tested using real-world \ac{EBC} observations and without scene information from conventional imaging sensors. By calibrating sources detected using the \ac{EBC}, a solution to the difficult challenge of converting raw event measurements into real-world measurements in a world coordinate frame is developed. In these speed survey experiments, the spatio-temporal characteristics of detected event sources are successfully related to the photometric characteristics of detected astrophysical objects. The results of this pipeline establish a foundation for principled space imaging and \ac{SSA} techniques using the \ac{EBC} with an improved understanding of the spatio-temporal features and measurement dynamics of event sources. In future work, several enhancements including improved star mapping, calibration and source finding algorithms are expected the further push the sensitivity limits and capabilities of the Gen 4 HD for space imaging and \ac{SSA}.

\section*{Conflict of Interest Statement}

The authors declare that the research was conducted in the absence of any commercial or financial relationships that could be construed as a potential conflict of interest.

\section*{Author Contributions}


NR conducted the investigation, developed the methodology and software, performed the validation, and wrote the original draft. AM collected the data used in this paper and reviewed the final draft. SA, NT, GC and AvS reviewed and edited the original draft, in addition to supervising the project and contributing to the methodology.

\section*{Funding}
This project was funded by Western Sydney University's Strategic Research Initiative. Some of the authors were supported by AFOSR grant FA9550-18-1-0471.

\ifCLASSOPTIONcaptionsoff
  \newpage
\fi



\bibliographystyle{IEEEtran}
\bibliography{main}

\end{document}